\documentclass[12pt]{article}

\usepackage{epstopdf,pdfpages,graphicx,lscape}
\usepackage{subfigure}
\usepackage[numbers,sort&compress]{natbib}
\usepackage{multirow,xcolor}
\usepackage{longtable, caption}
\usepackage{layout}
\usepackage{setspace} 
\usepackage[a4paper,bindingoffset=0.2in,%
            left=1in,right=1in,top=1in,bottom=1in,%
            footskip=.25in]{geometry}

\bibpunct[, ]{[}{]}{,}{n}{,}{,}

\title{Pharmacokinetics Simulations for Studying Correlates of Prevention Efficacy of Passive HIV-1 Antibody Prophylaxis in the Antibody Mediated Prevention (AMP) Study}

\author{Lily Zhang$^a$(yzhang2@fhcrc.org) \and Peter B. Gilbert$^{a,b}$(pgilbert@fhcrc.org) \and Edmund Capparelli$^c$(ecapparelli@ucsd.edu) \and Yunda Huang$^{a,d,\ast}$(yunda@fhcrc.org)}
\date{
$^a$Vaccine and Infectious Disease Division, Fred Hutchinson Cancer Research Center,  1100 Fairview Ave. N., Seattle, Washington, USA, 98109;
$^b$Department of Biostatistics, University of Washington, 1705 NE Pacific St., Seattle, Washington, USA, 98195;
$^c$Department of Pediatrics, University of California, San Diego, 9500 Gilman Drive \#0760, La Jolla, California, USA, 92093;
$^d$Department of Global Health, University of Washington, 1510 San Juan Rd., Seattle, Washington, USA, 98195; $^\ast$Corresponding author.}

\begin{document}
\maketitle

\newpage
\textbf{Abstract}
A key objective in two phase 2b AMP clinical trials of VRC01 is to evaluate whether drug concentration over time, as estimated by non-linear mixed effects pharmacokinetics (PK) models, is associated with HIV infection rate. We conducted a simulation study of marker sampling designs, and evaluated the effect of study adherence and sub-cohort sample size 
on PK model estimates in multiple-dose studies. With $m=120$, even under low adherence (about half of study visits missing per participant), reasonably unbiased and consistent estimates of most fixed and random effect terms were obtained. Coarsened marker sampling schedules were also studied. 


\textbf{keywords:} multiple-dose; NONMEM; population PK analysis; simulation-based sampling design; study adherence; two-compartment PK model.

\section{Introduction}
VRC01 is a human IgG1 broadly neutralizing monoclonal antibody (mAb) directed against the CD4-binding site of HIV-1 (\cite{WuMascola_Science2010, PeguNabelSCM2014,KoNabelNature2014, LedgerwoodGraham_CEI2015, MayerSeatonHuang2016}). 
In 2016, the HIV Vaccine Trials Network and HIV Prevention Trials Network launched the VRC01 Antibody Mediated Prevention (AMP) study, the first efficacy study of a broadly neutralizing anti-HIV antibody for prevention of HIV infection. The study is being conducted in two harmonized trials  in two cohorts: 2700 HIV-uninfected men and transgender persons in the Americas and Switzerland; and 1500 HIV-uninfected sexually active women in sub-Saharan Africa (\cite{GilbertJuraskaDeCamp2016}). Within each cohort, AMP participants are randomized to receive 10 IV infusions of 10 mg/kg VRC01, 30 mg/kg VRC01, or placebo every 8 weeks through 72 weeks. Besides the evaluation of prevention efficacy of VRC01, an important secondary objective of AMP is to assess VRC01 mAb markers, e.g. serum mAb concentrations over time, as correlates of protection (CoP) against HIV-1 infection. This knowledge is anticipated to help guide further development of mAbs and to provide benchmarks for HIV-1 vaccine development.\\

\noindent In the AMP correlates study, a case-control sampling design is used. Markers are measured from primary endpoint HIV-1 infected cases in the mAb groups and from a random sample of control participants from each mAb group who remain HIV-1 negative  until at least the Week 80 study visit  (\cite{GilbertJuraskaDeCamp2016}). While complete pharmacokinetics (PK) information (e.g. concentration at any given time) is desirable especially in the study of time-dependent correlates, it is infeasible to sample continuously or on an intensive time scale. Therefore, the full trajectory of VRC01 concentrations over time needs to be estimated based on the concentration data collected at selected time-points for the correlates study. As specified in the AMP protocol, 4-weekly blood samples were draw for possible marker measurements at up to 22 visits prior to Week 80  (i.e., 4 weeks and 8 weeks [trough] after each 8-weekly infusion), and at the visit scheduled 5 days after the second infusion (Figure 1). Of note, in the real trials these sampling times are subject to variations as a result of imperfect study adherence due to possible missed visits, permanent infusion discontinuations and study dropout, and actual visit windows around each target visit date (Supplementary materials). Therefore, besides different marker sampling designs, we also consider different levels of study adherence and their impact on the outcomes in the simulations described herein.\\

\noindent In this paper, we focus on the modeling of VRC01 concentrations among hypothetical HIV-uninfected control participants in AMP, defined as participants assigned to a VRC01 group who reach the Week 80 visit HIV-1 negative. Because all AMP participants acquiring HIV-1 infection during the study are sampled for measuring VRC01 concentrations at all available sampling time-points, the relevant sampling question is how many control participants and what time-points to sample in the AMP correlates study. Besides a range of sample sizes of control participants, we consider time-point sampling  according to: 1) the complete schedule of all 22 possible visits; and 2) coarsened schedules of only a subset of visits. In addition, because mAb serum concentrations often change non-linearly over time with possible individual-to-individual variability, population PK (popPK) analysis based on non-linear mixed effects models is used to estimate population and individual PK parameters, as well as to estimate individual-level serum concentrations over time (with associated estimation uncertainties) that can be assessed as potential correlates.\\

\noindent The outline of this paper is as follows. In Section 2, we introduce the Master PK model that is used for the simulation of AMP participants' serum concentration data; the simulation set-up, including different marker sampling designs for the AMP control cohort; and the method for estimating PK model parameters based on the simulated datasets. In Section 3, we compare the performance of different complete and coarsened schedule marker sampling designs in terms of the accuracy and precision of the PK parameter estimates. We draw some conclusions and recommendations in Section 4. 

\section{Methods}
\subsection{Master PK model}
A two-compartment model is deemed fitting to describe the PK of VRC01, since this mAb has limited distribution volume, slow clearance, and hence a long half-life ($\sim$15 days). Non-linear mixed effects models are used to describe individual-to-individual processes of drug absorption, distribution, and elimination and how these vary across individuals of different characteristics (e.g. weight, sex, or age). We use the popPK model that best describes the serum concentration data collected in a phase 1 study of VRC01 \cite{HuangZhangLedgerwood2017} as the prototype Master model for the simulation of concentration data from AMP participants. In the following, we briefly review the Master popPK model with two levels: the individual-level PK (or structure) model, which describes the intra-individual patterns of serum concentration over time through the processes of drug absorption, distribution, and elimination; and the population-level (or variability) model, which describes the inter-individual variability of the processes.

\subsubsection{Individual-level PK model}
 et $S(t_{ij})$ denote the VRC01 serum concentration at the $j^{th}$ measurement for subject $i$. In the Master individual-level PK model, it is defined as \vspace{-0.6cm}
$$S(t_{ij})= f(t_{ij},D_i,\beta_{i})*(1+e1_{ij}) + e2_{ij},\mbox{ for } i=1, \ldots, m, \mbox{ } j=1,\ldots,m_i$$ where $f$ denotes a two-compartment model with dose $D_i$, $\beta_i=(CL_i,V_{c_{i}},V_{p_{i}},Q_i)$ denotes parameters of $f$ specific to individual $i$, and the proportional error and additive constant error terms $e1_{ij}$ and $e2_{ij}$ satisfy, respectively, $\mbox{E}(e1_{ij}|D_i,\beta_i)=0$, $\mbox{var}(e1_{ij}|D_i,\beta_i)=\sigma_1^2$ and $\mbox{E}(e2_{ij}|D_i,\beta_i)=0$, $\mbox{var}(e2_{ij}|D_i,\beta_i)=\sigma_2^2$. There are different ways of parameterizing a compartment model. For their direct physiology significance, CL: clearance rate (L/day), $V_c$: volume of distribution in the central compartment (L), $V_p$: volume of distribution in the peripheral compartment (L), and Q: inter-compartmental clearance (L/day) are often a set of parameters used to describe PK processes underlying the observed concentration profiles for a given subject under a two-compartment model. \\

\noindent Specifically, after a single IV infusion, the serum drug concentration at time $t$ (relative to the end of infusion) can be  expressed as the sum of two exponential processes -- distribution and elimination (i.e. a two-compartment model):\vspace{-0.6cm}
\begin{eqnarray}
S(t) & = & \frac{D}{T_{inf}}\left[\frac{A}{\alpha}(1-e^{-\alpha T_{inf}})e^{-\alpha t} + \frac{B}{\beta}(1-e^{-\beta T_{inf}})e^{-\beta t}\right]\nonumber
\end{eqnarray}
where $D=$ IV dose amount (mg); $T_{inf}=$ duration of infusion; $\alpha$ and $\beta$ are rate constants (slopes) for the distribution phase and elimination phase, respectively ($\alpha > \beta$ by definition); and $A$ and $B$ are intercepts on the y axis for each exponential segment of the curve. Of note, when $\alpha T_{inf}\to 0$  and $\beta T_{inf} \to 0$ [as is the case for IV bolus administrations and is approximately the case for IV VRC01, where the infusion time is brief ($\sim$30 minutes) relative to the half-life], $S(t)=D[Ae^{-\alpha t} + Be^{-\beta t}]$ because $\lim_{h\to 0}\frac{e^h-1}{h}=1$. Let $k_{12}=Q/V_c$ and $k_{21}=Q/V_p$ represent the first-order rate transfer constants (1/day) for the movement of drug between the central and peripheral compartments. Mathematically, $\alpha$ and $\beta$ relate to the rate constants $k_{12}$ and $k_{21}$ as $\alpha+\beta  =  k_{12} + k_{21} + k$ and 
$\alpha \beta  =  k_{21}k$, where $k= \frac{CL}{V_c}= \frac{\alpha \beta(A+B)}{A\beta+B\alpha}$ is the rate of drug elimination. In addition,
$A =  \frac{\alpha-k_{21}}{V_c(\alpha-\beta)}$ and 
$B  =  \frac{k_{21}-\beta}{V_c(\alpha-\beta)} $.
After multiple doses, serum drug concentration at time $t$ after $k$ doses $D_j \mbox{ } (j=1,.\ldots, k)$ given at time $t_{D_j} (t \ge t_{D_k})$ can then be expressed as written below, based on the superimposed assumption that the PK of the drug after a single dose are not altered after taking multiple doses: \vspace{-0.6cm}
\begin{eqnarray}
S(t) &=& \sum_{j=1}^{k}D_{j}(Ae^{-\alpha(t-t_{D_j})} + Be^{-\beta(t-t_{D_j})}) \nonumber
\end{eqnarray}

\subsubsection{Population PK model}
Let $\beta_i = d(a_i, \beta, b_i), \mbox{ for } i=1, \ldots, m $
where $d$ denotes a $p$-dimensional function, $a_i$ denotes a vector of characteristics for subject $i$ that contribute to explaining inter-individual variability, $\beta$ denotes fixed effects, and $b_i$ denotes random effects. This equation characterizes how elements of $\beta_i$ vary across individuals due to systematic association with $a_i$ (modeled via $\beta$) and unexplained variation in the population represented by $b_i$. The usual assumptions $\mbox{E}(b_i|a_i) = \mbox{E}(b_i) = 0$ and $\mbox{var}(b_i|a_i)= \mbox{var}(bi) = \Omega$ are implied. \\

\noindent In the Master population-level PK model, all four individual-level PK parameters are modeled as log-normally distributed with exponential inter-individual variability (IIV) random effects. Let $CL_i$, $V_{c_i}$, $Q_i$, and $V_{p_i}$ denote individual-level estimated $CL$, $V_c$, $Q$, and $V_p$, respectively, for individual $i$. Covariates $a_i$ include only $BW_i$, body weight of subject $i$, which has an influence on $CL_i$ and $V_{c_i}$ via an exponential model as follows: 
\vspace{-0.6cm}
\begin{eqnarray}
CL_i &=& \beta_{CL} * \textrm{exp}( \beta_{BW.CL} * (BW_i-74.5)) * \textrm{exp}(b_{CL_i})\nonumber\\
V_{c_i} &=& \beta_{V_c} * \textrm{exp}( \beta_{BW.V_c} * (BW_i-74.5)) * \textrm{exp}(b_{V_{c_i}})\nonumber\\
Q_i &=& \beta_{Q} * \textrm{exp}(b_{Q_i})\nonumber\\
V_{p_i} &=& \beta_{V_p} * \textrm{exp}(b_{V_{p_i}})\nonumber
\end{eqnarray}
\vspace{-1cm}
\subsection{Simulation set-up}
\subsubsection{Study adherence}
Each participant's infusion and post-infusion visit schedule is simulated according to the AMP protocol specification (Supplementary Materials). Briefly, participants receive 10 IV infusions of VRC01 (or placebo). Each infusion visit has a visit window of -7 to +48 days around the scheduled 8-weekly infusion visit target date. Participants' marker measurements occur at each infusion visit, 4 weeks (visit window: -7 to +7 days) after each infusion, and 5 days (-2 to +2 days) after the second infusion. In the following simulations, for infusion visits, we assume that 80\% of the attended visits occur uniformly during the target window of -7 to +7 days and 20\% during the allowable window of +7 to +48 days. For post-infusion visits that occur between infusion visits, we assume all attended visits occur uniformly during the specified window. \\

\noindent Overall, we consider four study adherence patterns defined by the combinations of four missing data probabilities: probability of an independently missed single infusion ($p_1$), probability of an independently missed post-infusion visit ($p_2$), cumulative probability of permanent infusion discontinuation ($r_1$), and annual dropout rate ($r_2$). The four study adherence patterns in terms of ($p_1$, $p_2$, $r_1$, $r_2$) are: perfect adherence = (0\%, 0\%, 0\%, 0\%), high adherence = (5\%, 10\%, 10\%, 10\%), medium adherence = (10\%, 15\%, 15\%, 15\%), and low adherence = (15\%, 20\%, 20\%, 20\%). At 1 year after trial initiation, the adherence is very high in the ongoing AMP study with rates of  (2\%, 3\%, 5\%, 5\%). Nevertheless, a wider range of adherence levels are considered in this paper for interpolation purposes and for investigating the robustness of popPK modeling against missing data, relevant if adherence declines in AMP or for future mAb studies. 

\subsubsection{PK model parameter values}
Once participants' infusion and post-infusion visits are simulated according to the study adherence patterns described above, the Master popPK model described in section 2.1 is used to simulate VRC01 serum concentration at attended study visits for participants of a given body weight in each VRC01 dose group (10 mg/Kg and 30 mg/Kg). The PK parameter values used for the simulation are listed in Table \ref{Master PK Model}, where low covariances between random effects are fixed at zero to increase model stability. 

\subsubsection{Computation software}
R version 3.3.1 \cite{R2016} was used for the simulation of participants' characteristics and study visit data. The NONMEM software system (Version 7.3, ICON Development Solutions) was used for the simulation and modeling of concentration data. Parallel computing on eight central processing units was employed to speed up computation time via NONMEM. 

\subsubsection{Marker sampling design}
We consider two types of marker sampling designs: complete schedule and coarsened schedule (Figure 2). The former design includes concentration data at all time-points of visit attendance from each participant, whereas the latter includes concentration data only at a subset of time-points. Because a total of 61 VRC01 HIV-infected cases are expected at 60\% prevention efficacy for both dose groups pooled over both AMP trials, for the complete schedule design, we consider sample sizes of m = 30, 60, 120, and 240 HIV-1 uninfected controls. The latter three sample sizes correspond to the numbers of expected controls in the case-control cohort with a 1:1, 1:2, and 1:4 case:control ratio, respectively, whereas m = 30 serves as a reference and represents the 1:1 case:control ratio for a single AMP trial. For the coarsened schedule design, we consider only one sample size of m=240, but 3 coarsened schedules that sample roughly half of the complete schedule time-points per participant. 
\begin{itemize}
\item First half: the first 11 time-points (excluding time 0) out of the total 22 complete schedule time-points are sampled. These time-points include, for every individual the 4-week and 8-week post-infusion time-points (trough) after the first 5 infusions, in addition to the 5-day time-point after the 2nd infusion. 
\item Mixed half: time-points after every other infusion are sampled. In addition to the 5-day time-point after the $2^{nd}$ infusion for every individual, 4-week and 8-week time-points (trough) after the five odd number ($1^{st}$, $3^{rd}$, $5^{th}$, $7^{th}$, and $9^{th}$) infusions are included for one half of the subjects ($m=120)$, and 4-week and 8-week post-infusion time-points after the five even number ($2^{nd}$, $4^{th}$, $6^{th}$, $8^{th}$, and $10^{th}$) infusions are included for the other half of the subjects ($m=120$).
\item Trough only: trough time-points are sampled. These include, for every individual the 8-week time-points (trough) after each of the 10 infusions, in addition to the 5-day time-point after the 2nd infusion.
\end{itemize}
\noindent Note that for all 3 coarsened schedules, the 5-day time-point after the $2^{nd}$ infusion is always included because this  time-point is the only one scheduled within a few days of an infusion for the estimation of $V_c$. These 3 coarsened schedules are compared to the complete schedule with $m=120$, since the same number of 4-week and 8-week post infusion observations are made. 

\subsubsection{Simulation steps}
For the complete schedule marker sampling design, we consider a total of 16 scenarios representing combinations of 4 study adherence patterns (perfect, high, medium, and low) and 4 sample sizes (m=30, 60, 120, and 240). For the coarsened schedule marker design, we consider a total of 12 scenarios representing the combinations of 4 study adherence patterns and 3 coarsened schedules (First half, Mixed half, and Trough only). The complete schedule datasets with m=240 are first simulated and the coarsened schedule datasets are extracted from them given the specific design. For each scenario, 1000 datasets are simulated containing participants' demographic information (body weight and sex), study information (infusion and post-infusion visit time), and serum concentration at attended visits. Depending on the research interest, concentration values on consecutive days or flexible grid can also be simulated to evaluate the modeling/prediction of simulated concentration using sparse data. Specifically, each dataset is simulated following the steps described below:
\begin{itemize}
\item[Step 1:] Simulate participants' characteristics representing the two AMP study cohorts and their corresponding VRC01 dose amounts in the low and high dose groups. 
\begin{itemize}
\item[a.]Simulate 1:1 male:female sex ratio.
\item[b.]For females and males, respectively, randomly sample participants' body weight with replacement from females in the Phambili trial in South Africa \cite{GrayMoodieMetch2014}, and from males in the Step study in the Americas \cite{BuchbinderMehrotraDuerr2008}. Participants' body weights are assumed to be constant throughout the trial. 
\item[c.] Within each sex, participants are randomly assigned to the 10 mg/Kg and 30 mg/Kg dose groups at a 1:1 ratio. Each participant's VRC01 dose amount is determined as the product of his/her body weight and dose level.
\end{itemize}
\vspace{0.2cm}
\item[Step 2:] Simulate participants' 8-weekly infusion visits.
\begin{itemize}
\item[a.] Simulate the attendance (yes or no) of each of the 10 infusion visits using a Bernoulli probability of $p_1$.
\item[b.] If attendance is `yes' for a given infusion visit, simulate the infusion visit time according to the AMP protocol-specified schedules and visit windows, assuming the probability of attending the visit within the target window (typically -7 to +7 days) is 80\% and the probability of attending the visit outside the target window but within the study allowable window (typically -7 to +48 days) is 20\%. The infusion time follows a piece-wise uniform distribution within and outside the target window.
\end{itemize}
\vspace{0.2cm}
\item[Step 3:] Simulate participants' post-infusion visits: baseline, 4-weekly, and 5-day post-2$^{nd}$ infusion.
\begin{itemize}
\item[a.] Simulate post-infusion visit attendance (yes or no) according to prior infusion attendance. If the prior infusion is administered, simulate the visit attendance using a Bernoulli probability of $p_2$. If the prior infusion is missed, then the 5-day post-2$^{nd}$ infusion visit (if applicable) and the following 4-week post-infusion visits are all considered missed. If the last infusion (\#10) is missed, then the 4-week post-infusion visit is also missed. However, the 8-week post infusion visit could still be scheduled.
\item[b.] If post-infusion attendance is `yes', simulate post-infusion visit time accounting for prior infusion visit time, according to protocol-specified schedules/windows, similar to the procedure described in Step 2b. 
\end{itemize}
\vspace{0.2cm}
\item[Step 4:] Modify infusion attendance and post-infusion visits accounting for permanent infusion discontinuation.
\begin{itemize}
\item Simulate time to permanent infusion discontinuation (due to reasons other than HIV infection) using a random exponent rate $r_1$.
\item Modify infusion attendance and post-infusion visits for those who discontinue infusion permanently according to a different protocol-specified visit schedule/window (Supplementary Materials).
\end{itemize}
\vspace{0.2cm}
\item[Step 5:] Modify infusion attendance and post-infusion visits accounting for dropout (study termination).
\begin{itemize}
\item Simulate time to dropout using a random exponent rate $r_2$.
\item Censor all previously simulated infusion and post-infusion visits at dropout time.
\end{itemize}
\vspace{0.2cm}
\item[Step 6:] Simulate participants' concentrations according to the Master PK model, with the covariate information, dose amount, infusion, and visit schedules as in the previous steps. 
\begin{itemize}
\item[(a)] Simulate $S(t_{ij})$ according to the final PK model by setting $b_i=0$ and $\epsilon_{ij}=0$.
\item[(b)] Add the above value to a mean-zero normally distributed $b_i$ and $\epsilon_{ij}$, according to the variance estimates from the final PK model.
\end{itemize}
\end{itemize}

\subsection{Estimation Method} 
Parameter estimation of the popPK model is based on minimizing the objective function value using maximum likelihood estimation. A marginal likelihood of the observed data is calculated based on both the influence of the fixed effect and the random effect. Different estimation methods for non-linear mixed effects models have been extensively discussed by other authors (e.g. \cite{Plan_AAPS2012, Johansson_JPP2014}). In this paper, due to the sparseness of the simulated data, the Markov chain Monte Carlo stochastic approximation expectation-maximization (SAEM) method \cite{Delyon_AS1999,Kuhn_ESAIM2004} is applied to the modeling of the simulated time-concentration data according to the Master PK model. The true values of each PK parameters as specified in Table 1 are used as initial values. 

\section{Results}
For each scenario of the complete schedule and coarsened schedule designs, we report 
\begin{itemize}
    \item[1)] \% datasets `converged'. Due to the Monte Carlo nature of the SAEM method, convergence testing is not formally done.  Completed runs are counted as convergence successes.
    \item[2)] For each fixed and random effect: 6 fixed-effect terms, 6 random-effect terms, and 2 residual error terms, among the B converged models, 
    \begin{itemize}
    \item relative bias, RBias $ =\frac{1}{K}\sum^{B}_{k=1}(\frac{\hat\beta_k - \beta}{\beta})*100$, 
    \item relative root mean squared error, RRMSE $=\sqrt[]{\frac{1}{K}\sum^{B}_{k=1}(\frac{\hat\beta_k - \beta}{\beta})^{2}}*100$, and 
    \item coverage probability, CP$=$proportion of datasets with 95\% confidence intervals including the true value of the parameter $\beta$. 
\end{itemize}
\item[3)] Shrinkage estimate for each random-effect terms, i.e. one minus the ratio of the standard deviation of the individual-level estimate and the estimated variability for the population estimate. 
\end{itemize}

\subsection{Master popPK model}
As an illustration, Figure 2 displays two expected population-level time-concentration curves for individuals who are perfectly adherent to the 8-weekly infusion schedule, one for the 10 mg/Kg dose and one for the 30 mg/Kg dose, based on the master popPK model with $E(b_i)=0$, $E(e1)=0$, and $E(e2)=0$. The concentrations at each time-point are simulated based on a body weight of 74.5 Kg and with the PK parameter values given in Table 1.

\noindent In addition, a random set of the simulated individual-level concentration curves under low study adherence are displayed in Figure 3. For example, individual \#2 in the low dose group (left panel) stayed in the study for follow up but discontinued infusions after the first infusion, whereas individuals \#9 and \#26 in the high dose group (right panel) dropped out of the study right after the first post-infusion visit.

\subsection{Model fitting}
\subsubsection{Complete schedule marker sampling designs}
Using the SAEM estimation method, almost all models using datasets under the complete schedule designs converged to obtain final PK parameter estimates for $m=120$ and $m=240$, whereas a relative low convergence was observed for datasets with $m=30$ and $m=60$ (Figure 4). This result suggests that a minimal sample size of $m=60$ is recommended for a stable PK model fitting under the described schedule. This suggestion is further confirmed when the accuracy and precision of the fixed-effect estimates are examined (Figures 5 \&6, Supplementary Materials: Table 1). Reasonable levels of bias and precision with significant improvements over the small sample sizes are observed for $m=120$. On the other hand, $m=240$ provides relatively marginal improvements compared to $m=120$ except for $\beta_{BW.CL}$, which characterizes the effect of each individual's body weight on CL and requires data from a sufficient number of independent subjects for an accurate and precise estimation.\\

\noindent Due to the unstable estimation for $n=30$, we restrict the evaluation of the estimation of random effects to larger sample sizes (Supplementary Materials: Figures S1 \& S2). In general, regardless of the sample size, the random effect of the PK parameter Q (inter-compartmental clearance rate) is poorly estimated with high shrinkage (Supplementary Materials: Figure S3). This poor estimation is due to the sparsity of data closer to infusion, with only one 5-day post infusion time-point, and the low inter-individual variability in Q. On the other hand, the estimation of CL and $V_p$ seems reasonable, with shrinkage generally below 20-30\%. The proportional error term, $\sigma_1^2$ is also reasonably estimated with RRMSE $< 10\%$ under all scenarios. The estimation of the additive error term, $\sigma_2^2$ is relatively poorer possibly due to the sparsity of data around the assay limit of detection. 

\subsubsection{Coarsened schedule marker sampling designs}
Figure 7 displays the distribution of the total number of observations (5 days after the $2^{nd}$ infusion, 4 weeks and 8 weeks after each infusion) based on the 1000 simulated datasets under each of the 12 coarsened schedule scenarios with m=240, along with the complete schedule design with m=120. Because about half of the complete schedule time-points are sampled in the coarsened schedules, the total expected number of 4-week and 8-week post infusion observations are the same across the 4 designs. This feature allows a fair comparison across the designs. Meanwhile, the number of 5-day post second infusion observations is doubled in the coarsened schedule designs, because the 5-day post $2^{nd}$ infusion time-point is the only time-point proximal to an infusion. Hence, this time-point is always sampled from every individual under both the complete and coarsened schedule designs. This consistency allows the assessment of the impact of 5-day post infusion observations on the estimation of various PK parameters.\\

\noindent Results showing the accuracy and precision of the fixed-effect estimates under each coarsened schedule design with $m=240$ are displayed in Figures 8, 9 and Supplemental Materials (Table 2). The complete schedule design with $m=120$ is included as a reference for comparison purposes. In general, for the estimation of fixed effects, the complete schedule design with half of the sample size provides more accurate but less precise estimates, except that more accurate estimates are obtained for $\beta_{V_c}$ and $\beta_{BW.CL}$ under the `First half' and `Mixed half' coarsened schedule designs. This is likely due to the fact that having more 5-day post infusion observations in the coarsened schedule designs helps improve the estimation of $\beta_{V_c}$, which requires data proximal to infusion for an accurate estimation, and having more independent individuals helps improve the estimation of the covariate effect. On the other hand, the accuracy of estimates under the coarsened schedule designs are more impacted by study adherence due to the sparser time-points compared to the complete schedule design. Among the 3 coarsened schedule designs, the `First half' and `Mixed half' designs have very similar performance and are generally superior to the `Trough only' design, especially for the estimation of $\beta_{V_c}$. Similar patterns are observed in the estimation of random effects (Supplemental Materials: Figures S4-S6). Poor estimation of the random effect of inter-compartment clearance (Q) and additive residual error are observed for all designs for reasons stated above.

\section{Conclusions}
PopPK analysis is known to be suitable for datasets consisting of a few data points per individual over the course of product administration(s) from many individuals, in order to estimate popPK parameters adjusting for variability among individuals. In this paper, we investigated how the accuracy and precision of the estimated population parameters (fixed effects) and variabilities among individuals (random effects) are influenced by the number of individuals and by the number and type (i.e. time-point) of observations per individual.\\

\noindent In the context of the AMP study, where participants receive ten 8-weekly IV infusions of VRC01, we considered complete schedule marker sampling designs where approximately 4-weekly observations from up to 22 time-points over the course of 80 weeks are included in the popPK modeling, with 4 different levels of study adherence (perfect, high, medium, and low) and 4 different sample sizes (m= 30, 60, 120, and 240). We found that a sample size of 120 or higher could render reasonably unbiased and consistent estimates of most fixed and random effect terms. The central volume parameter $V_c$ is the most challenging fixed effect parameter to estimate due to the lack of concentration data proximal to infusion as specified in the AMP protocol.\\ 

\noindent We also considered coarsened schedule marker sampling designs with $m=240$, where the first half (`First half'), alternate (`Mixed half'), or `Trough only' time-points are included in the popPK modeling. These designs often provide less accurate but more precise estimates of various popPK parameters than the complete schedule design with $m=120$. In terms of overall estimation performance as measured by RRMSE, the `First half' and `Mixed half' designs render similar performance, but are generally superior to the `Trough only' design and the complete schedule design. We note that the `First half' design is less subject to missing data, but provides limited data in the assessment of the steady state and the effect of a higher number of repeated doses. On the other hand, the `Mixed half' design is more subject to missing data due to infusion discontinuation and study drop out as the study progresses. Based on these simulation results, we favor using the `Mixed half' design for the AMP case-control study, given that it provides the best overall accuracy and precision for various PK parameter estimates in studies of high adherence like the current AMP study. In addition, the `Mixed half' design allows the assessment of concentrations after any of the ten infusions (as opposed to only the first five infusions in the `First half' design); these data may be helpful in the analysis and interpretation of other study endpoints including long-term safety and anti-drug activity that may occur later in the study. If adherence declines in AMP, the advantages of the coarsened schedule designs will diminish and the full schedule design may be considered. \\

\noindent In summary, this paper provides a simulation-based framework to evaluate sampling designs of multiple-dose PK studies using a stochastic process for participants' characteristics (e.g. sex and body weight) and infusion/measurement time-points. It also provides a simulator for studying statistical methods for assessing prevention efficacy and correlates of prevention efficacy. This simulator not only accounts for participant characteristics that influence PK processes, but also accounts for possible missed or terminated product administrations, protocol-specific study visits and visit windows, and potential drop out. Thus, this framework provides a realistic simulator of PK data for future studies of repeatedly-administered drugs. 

\section*{Disclosure statement}
No potential conflicts of interest were disclosed. 
\section*{Funding}
This work was supported by the National Institute of Allergy and Infectious Diseases (NIAID) US. Public Health Service Grant UM1 AI068635 [HVTN SDMC FHCRC]. The content of this manuscript is solely the responsibility of the authors and does
not necessarily represent the official views of the National Institutes of Health. The funders had no role in study design, data collection and analysis, decision to publish, or preparation of the manuscript.

\newpage


\begin{thebibliography}{}
\bibitem{WuMascola_Science2010}
Wu X, Yang ZY, Li Y, et al. Rational design of envelope identifies broadly neutralizing human monoclonal antibodies to HIV-1. Science. 2010;329(5993):856-61.

\bibitem{PeguNabelSCM2014}
Pegu A, Yang ZY, Boyington JC, et al. Neutralizing antibodies to HIV-1 envelope protect more effectively in vivo than those to the CD4 receptor. Science Translational Medicine. 2014;6(243):243ra88.

\bibitem{KoNabelNature2014}
Ko SY, Pegu A, Rudicell RS,et al. Enhanced neonatal Fc receptor function improves protection against primate SHIV infection. Nature. 2014;514(7524):642-5.

\bibitem{LedgerwoodGraham_CEI2015}
Ledgerwood JE, Coates EE, Yamshchikov G,et al. Safety, pharmacokinetics and neutralization of the broadly neutralizing HIV-1 human monoclonal antibody VRC01 in healthy adults. Clinical and Experimental Immunology. 2015;182(3):289-301.

\bibitem{MayerSeatonHuang2016}
Mayer KH, Seaton K, Huang Y, et al. Clinical safety and pharmacokinetics of IV and SC VRC01, a broadly neutralizing mAb. Abstract presented at: CROI 2016; Boston, USA.

\bibitem{GilbertJuraskaDeCamp2016}
Gilbert PB, Juraska M, deCamp AC, et al. Basis and Statistical Design of the Passive HIV-1 Antibody Mediated Prevention (AMP) Test-of-Concept Efficacy Trials. Statistical Communications in Infectious Diseases. In Press.

\bibitem{HuangZhangLedgerwood2017}
Huang Y, Zhang Y, Ledgerwood JE, et al. Population pharmacokinetics analysis of VRC01, an HIV-1 broadly neutralizing monoclonal antibody, in healthy adults. mAbs. 2017 Apr 3:0. doi: 10.1080/19420862.2017.1311435. [Epub ahead of print]

\bibitem{R2016} 
R Core Team (2016). R: A language and environment for statistical computing. R Foundation for Statistical Computing, Vienna, Austria. URL https://www.R-project.org/.

\bibitem{EMA_2014} 
EMA. Guideline on the investigation of bioequivalence. CPMP/EWP/QWP/1401/98 Rev. 1/ Corr **. London, 20 January 2010.

\bibitem{GrayMoodieMetch2014}
Gray GE, Moodie Z, Metch B, et al. Recombinant adenovirus type 5 HIV gag/pol/nef vaccine in South Africa: unblinded, long-term follow-up of the phase 2b HVTN 503/Phambili study. Lancet Infect Dis. 2014;14(5):388-96.

\bibitem{BuchbinderMehrotraDuerr2008}
Buchbinder SP, Mehrotra DV, Duerr A, et al. Efficacy assessment of a cell-mediated immunity HIV-1 vaccine (the Step Study): a double-blind, randomised, placebo-controlled, test-of-concept trial. Lancet. 2008;372(9653):1881-93.

\bibitem{LindstromBates_Biometrics1990} 
Lindstrom LM, Bates DM. Nonlinear Mixed-Effects Models for Repeated Measures Data. Biometrics. 1990;46:673-687.
 
\bibitem{Plan_AAPS2012} 
Plan EL, Maloney A, Mentre F, et al. Performance Comparison of Various Maximum Likelihood Nonlinear Mixed-Effects Estimation Methods for Dose-Response Models. The American Association of Pharmaceutical Scientists Journal. 2012;14(3):420-32.

\bibitem{Johansson_JPP2014} 
Johansson MA, Ueckert S, Plan E, et al. Evaluation of bias, precision, robustness and runtime for estimation methods in NONMEM 7. J Pharmacokinet Pharmacodyn. 2014;41:223-38.


\bibitem{Delyon_AS1999} 
Delyon B, Lavielle M, Moulines E.  Convergence of a stochastic approximation version of the EM algorithm. Ann. Statist. 1999;27(1):94-128.

\bibitem{Kuhn_ESAIM2004} 
Kuhn E, Lavielle M. Coupling a stochastic approximation version of EM with an MCMC procedure. ESAIM: P\&S 2004;8:115-131.

\end{thebibliography}
\newpage
\begin{landscape}

\begin{center}
\begin{table}[thp]
\caption{\bf Master popPK model parameter values for the simulation of VRC01 time-concentration data.\\}
\label{Master PK Model}
\begin{tabular}{lccc}
\hline
\hline 
Term & Parameter & True value & Note\\
\hline
\hline 
\multirow{6}{*}{Fixed effects} & $\beta_{CL}$ &0.41 & Clearance from the central compartment (CL: L/day)\\
& $\beta_{V_c}$ & 1.92 & Central compartment volume ($V_c$: L)\\
& $\beta_{Q}$ & 0.87 & Inter-compartment clearance (Q: L/day)\\
& $\beta_{V_p}$ & 5.32 & Peripheral compartment volume ($V_p$: L)\\
& $\beta_{BW.CL}$ & 0.0072 & Body weight influence on CL (/Kg)\\
& $\beta_{BW.V_c}$ & 0.011 & Body weight influence on $V_c$ (/Kg)\\
\hline
                            & $\mbox{var}(b_{CL})$& 0.075 & Inter-individual variability of CL\\
Variance of & $\mbox{var}(b_{V_c})$& 0 &  Inter-individual variability of $V_c$. Fixed at zero.\\
random effects              & $\mbox{var}(b_{Q})$& 0.091 &Inter-individual variability of Q\\
                            & $\mbox{var}(b_{V_p})$& 0.14 &Inter-individual variability of $V_p$\\
\hline
& $\mbox{cov}(b_{CL}, b_{V_c})$ & 0 & Fixed at zero\\
& $\mbox{cov}(b_{CL}, b_{Q})$ & 0.05 & \\
Co-variance of & $\mbox{cov}(b_{CL}, b_{V_p})$ & 0.062&\\
random effects & $\mbox{cov}(b_{V_c}, b_{Q})$ & 0 & Fixed at zero\\
& $\mbox{cov}(b_{Vc}, b_{V_p})$ & 0  & Fixed at zero\\
& $\mbox{cov}(b_{Q}, b_{V_p})$ & 0.11 &\\
\hline
Variance of & $\sigma^2_1$ & 0.044 & Proportional error\\
residual errors & $\sigma^2_2$ & 0.43 & Additive constant error\\
\hline
\end{tabular}
\addtocounter{table}{-1} 
\end{table}
\end{center}
\end{landscape}
\clearpage

\textbf{Figure Legends}\\

\textbf {Figure 1: AMP study schema and marker sampling designs.} Panel A: a total of 2700 participants from the Americas and Switzerland trial and 1500 participants from the sub-Saharan Africa trial receive ten IV infusions (\#1- \#10) of VRC01 at 10 mg/Kg, 30 mg/Kg or placebo every 8-weekly at a 1:1:1 randomization ratio. Arrows indicate pre- and post-infusion time-points included in the complete schedule marker sampling design. Panel B: Orange circles indicate sampled time-points in each coarsened schedule marker sampling design. Additional specimen collection times at baseline (week 0) and after week 80 are not included in this figure because data at these time-points mainly contribute to assay control and safety monitoring, not in the PK modeling.\\  
\begin{center}
\includegraphics[width=1.0\textwidth]{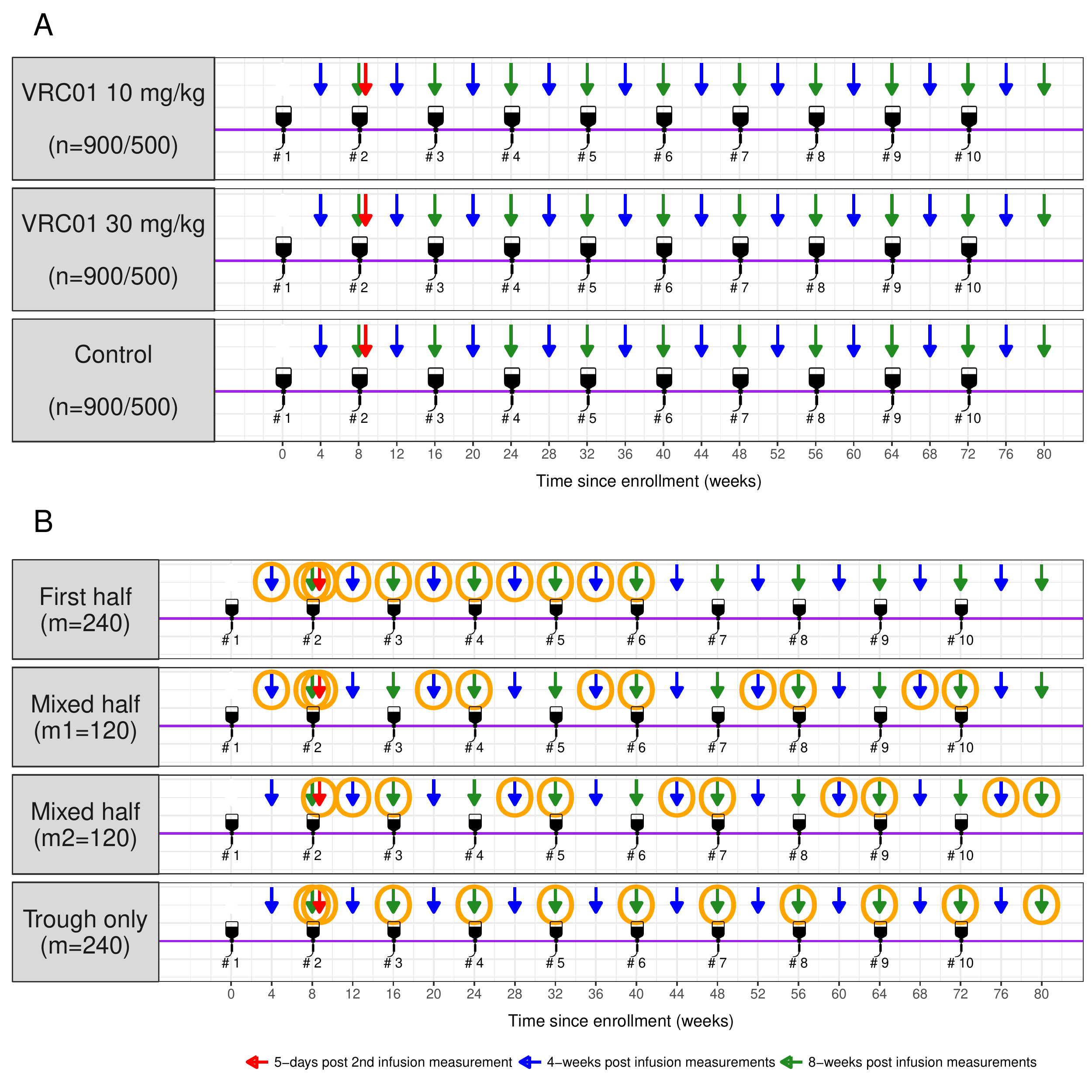}
\end{center}

\clearpage
\textbf {Figure 2: Simulated time-concentration curves under the Master popPK model following ten 8-weekly IV infusions of VRC01 in the 10 mg/Kg and 30 mg/Kg dose groups with perfect study adherence.} Solid lines are the medians; shaded areas are the 2.5\% and 97.5\% percentiles of the concentrations over 1000 simulated datasets. A body weight of 74.5 Kg is used in the simulations.\\
\begin{center}
\includegraphics[width=1.0\textwidth]{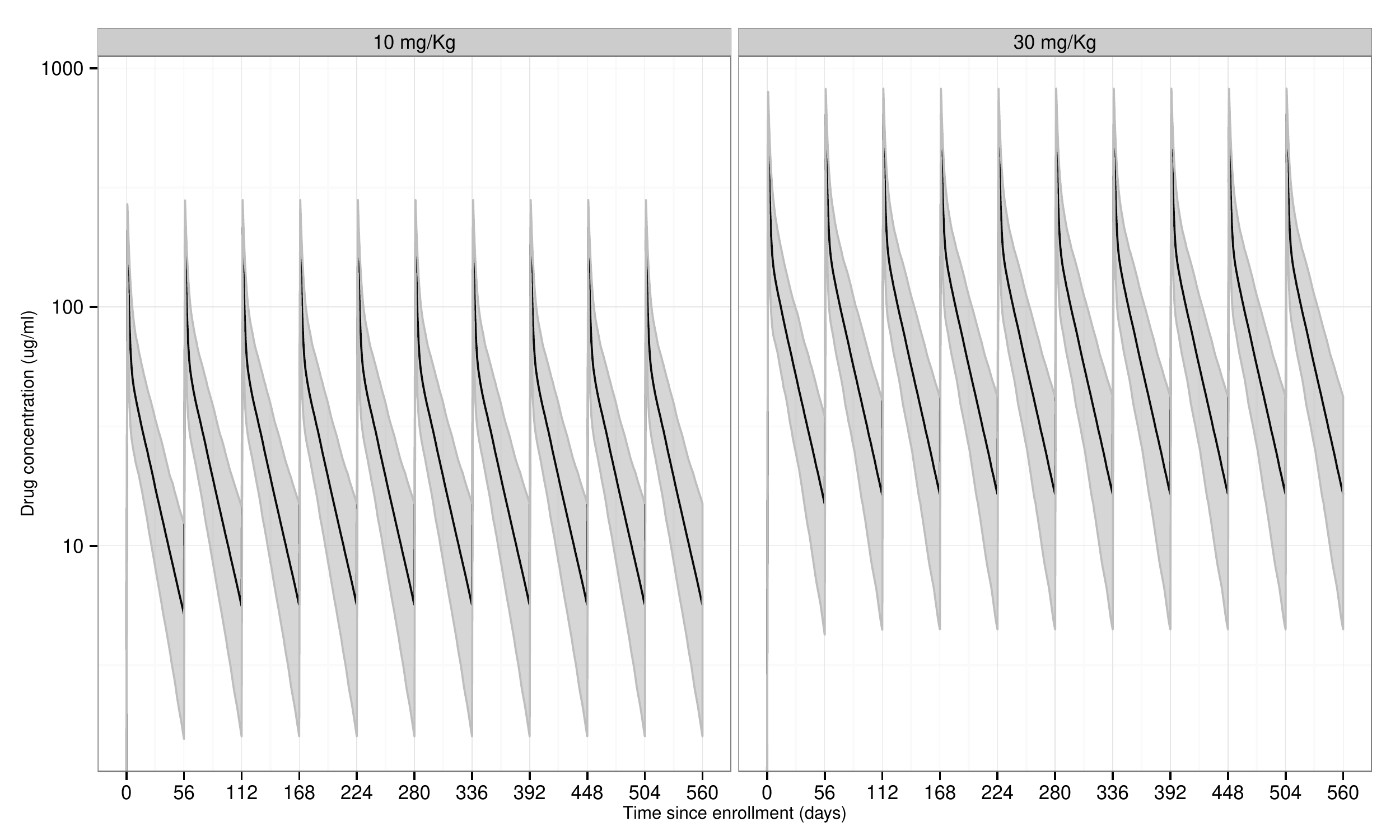}
\end{center}

\newpage
\textbf {Figure 3: Simulated time-concentration curves under the Master popPK model following ten 8-weekly IV infusions of VRC01 in the 10 mg/Kg and 30 mg/Kg dose groups with low study adherence.} With low study adherence, probability of an independently missed single infusion ($p_1$), probability of an independently missed post-infusion visit ($p_2$), cumulative probability of permanent infusion discontinuation ($r_1$), and annual drop out rate ($r_2$) are assumed to be 15\%, 20\%, 20\%, and 20\%, respectively.\\ 
\begin{center}
\includegraphics[width=1.0\textwidth]{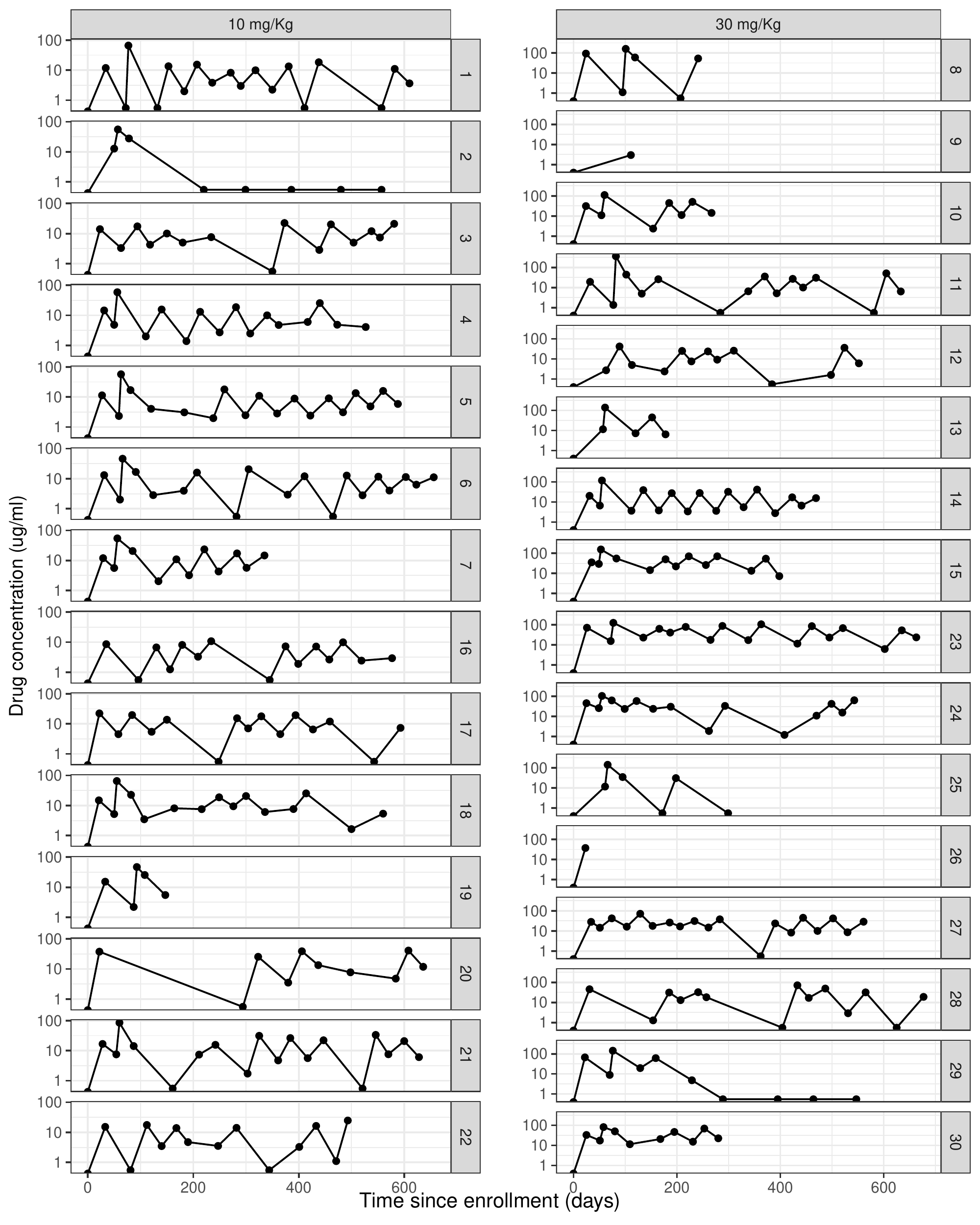}
\end{center}

\newpage
\textbf {Figure 4: Percent of models converged under the complete schedule designs with different levels of study adherence and sample sizes (m = 30, 60, 120, and 240.)} The four study adherence patterns in terms of ($p_1$, $p_2$, $r_1$, $r_2$) are: Perfect = (0\%, 0\%, 0\%, 0\%), High = (5\%, 10\%, 10\%, 10\%), Medium = (10\%, 15\%, 15\%, 15\%), and Low = (15\%, 20\%, 20\%, 20\%), where $p_1$ indicates the probability of an independently missed single infusion, $p_2$ the probability of an independently missed post-infusion visit, $r_1$ the cumulative probability of permanent infusion discontinuation, and $r_2$ the annual drop out rate.\\ 
\begin{center}
\includegraphics[width=1.0\textwidth]{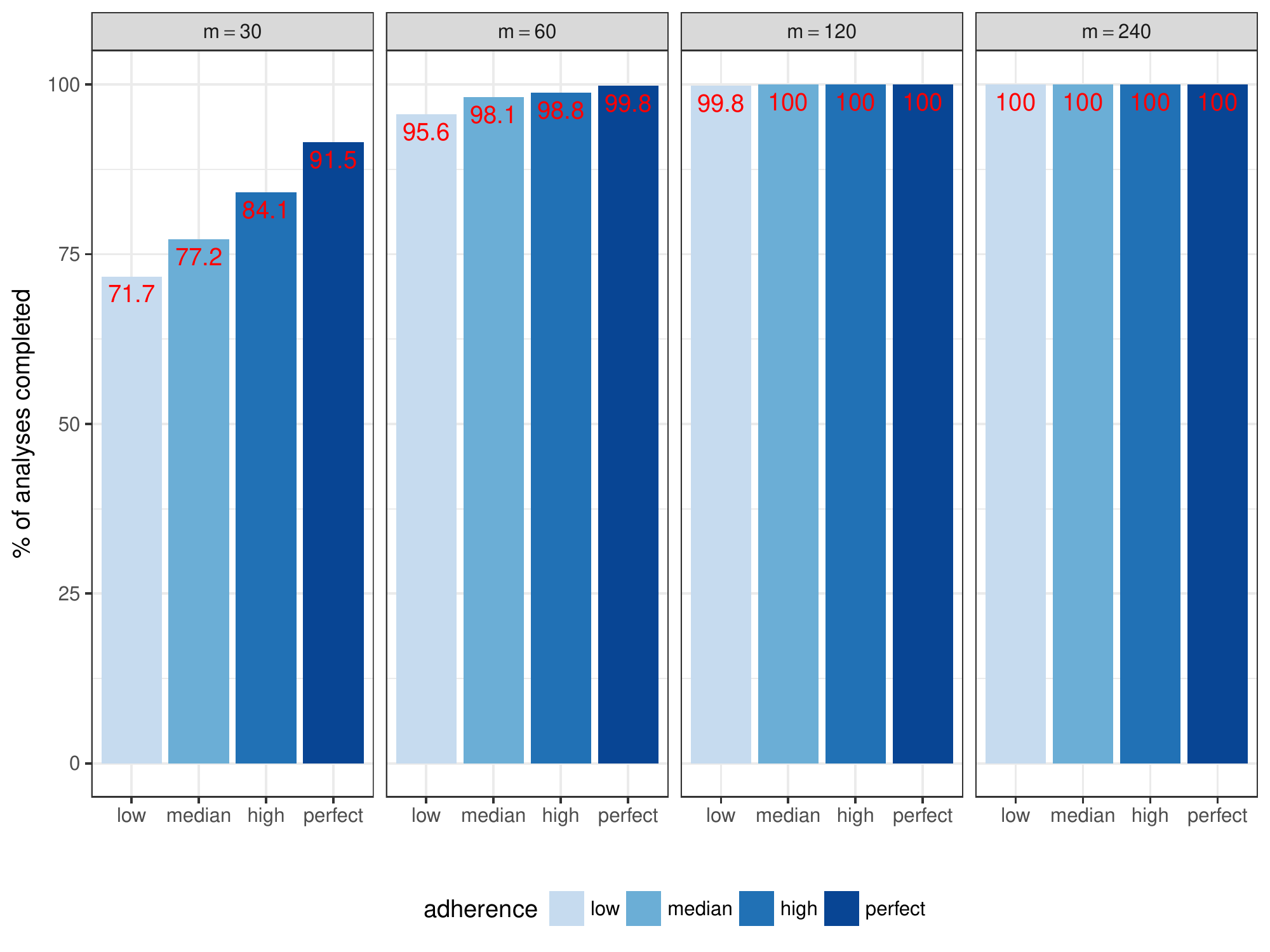}
\end{center}

\newpage
\textbf {Figure 5: Relative bias of each fixed-effect PK parameter estimate under complete schedule designs with different sample sizes (m = 30, 60, 120, and 240).} $\beta_{CL}$, $\beta_{V_c}$, $\beta_{Q}$, and $\beta_{V_p}$ indicate the fixed effect for CL, $V_c$, Q, and $V_p$, respectively. $\beta_{BW.CL}$ and $\beta_{BW.V_c}$ indicate the fixed effect of body weight influence on CL and $V_c$, respectively.\\
\begin{center}
\includegraphics[width=1.0\textwidth]{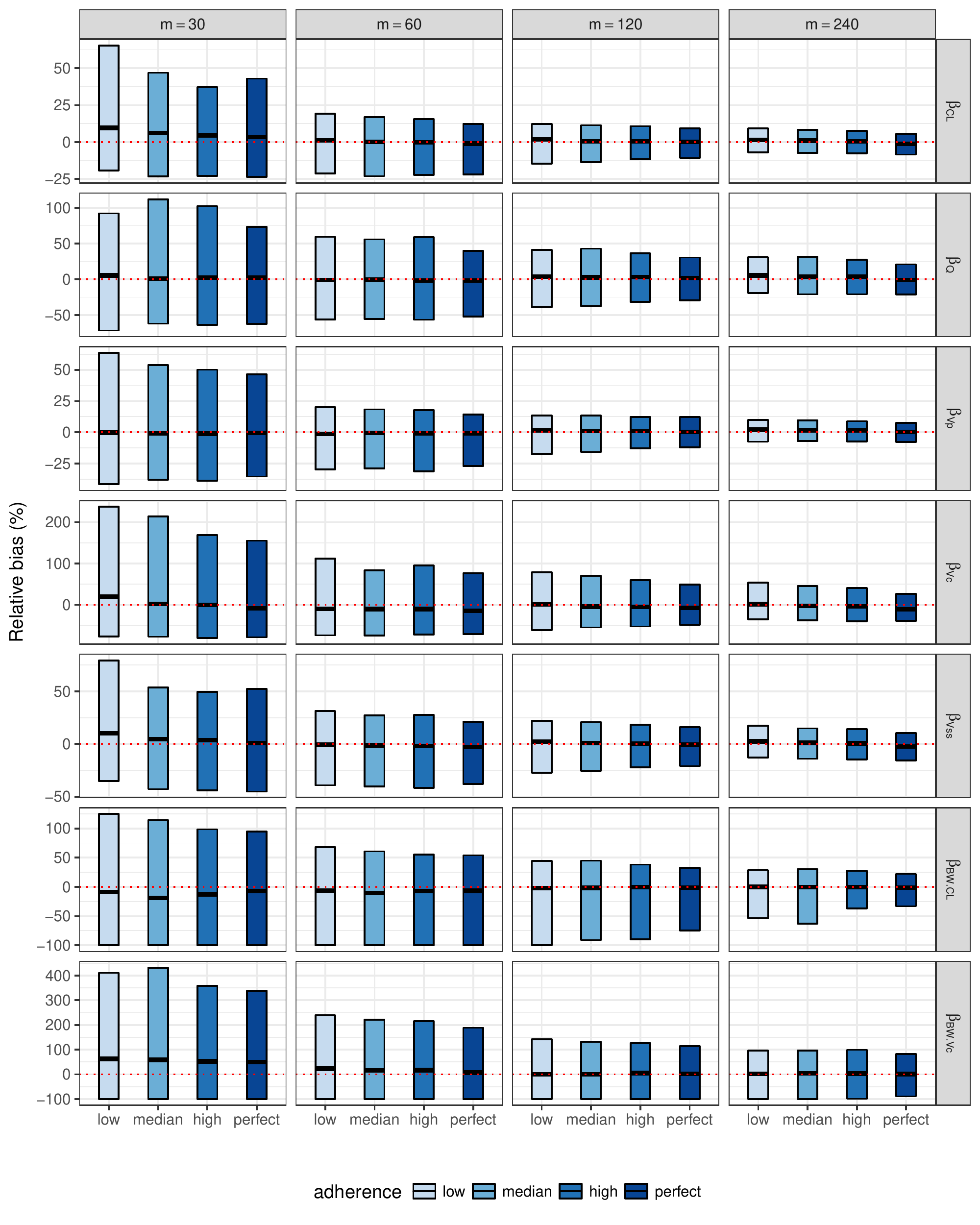}
\end{center}

\newpage
\textbf {Figure 6: Relative root mean squared errors of each fixed-effect PK parameter estimate under complete schedule designs with different sample sizes (m = 30, 60, 120, and 240).} $\beta_{CL}$, $\beta_{V_c}$, $\beta_{Q}$, and $\beta_{V_p}$ indicate the fixed effect for CL, $V_c$, Q, and $V_p$, respectively. $\beta_{BW.CL}$ and $\beta_{BW.V_c}$ indicate the fixed effect of body weight influence on CL and $V_c$, respectively.\\
\begin{center}
\includegraphics[width=1.0\textwidth]{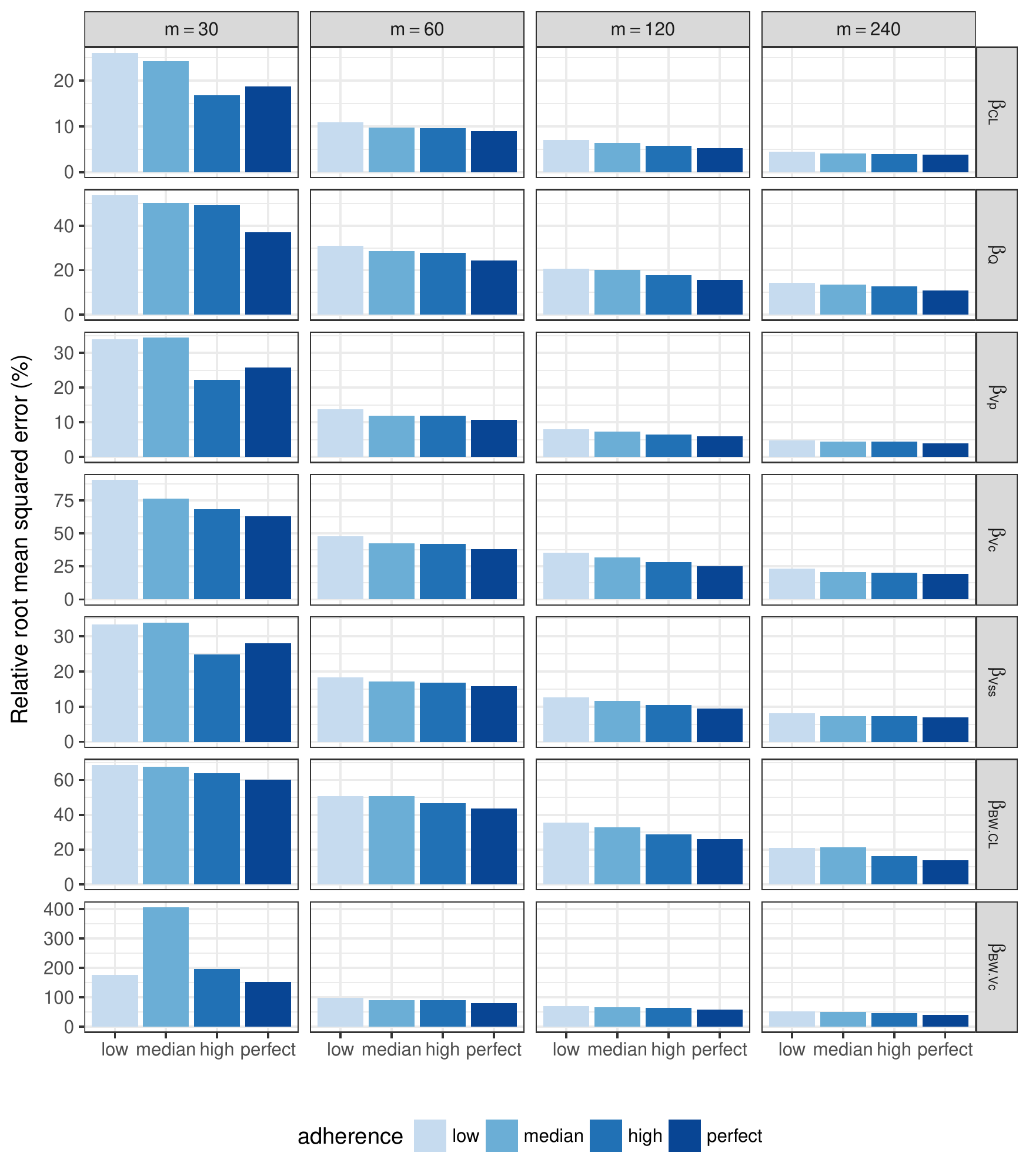}
\end{center}

\newpage
\textbf {Figure 7: Distributions of the total numbers of 5-day post $2^{nd}$ infusion, 4-week post infusion, and 8-week post infusion observations under the complete schedule design with m = 120 and 3 coarsened schedule designs with m = 240.} The `First half' design samples the first 11 time-points (excluding time 0) out of the total 22 complete schedule time-points. The `Mixed half' design samples time-points after every other infusion. The `Trough only' design samples only trough time-points. All 3 coarsened schedule designs always include the 5-day post $2^{nd}$ infusion time-point.\\
\begin{center}
\includegraphics[width=1.0\textwidth]{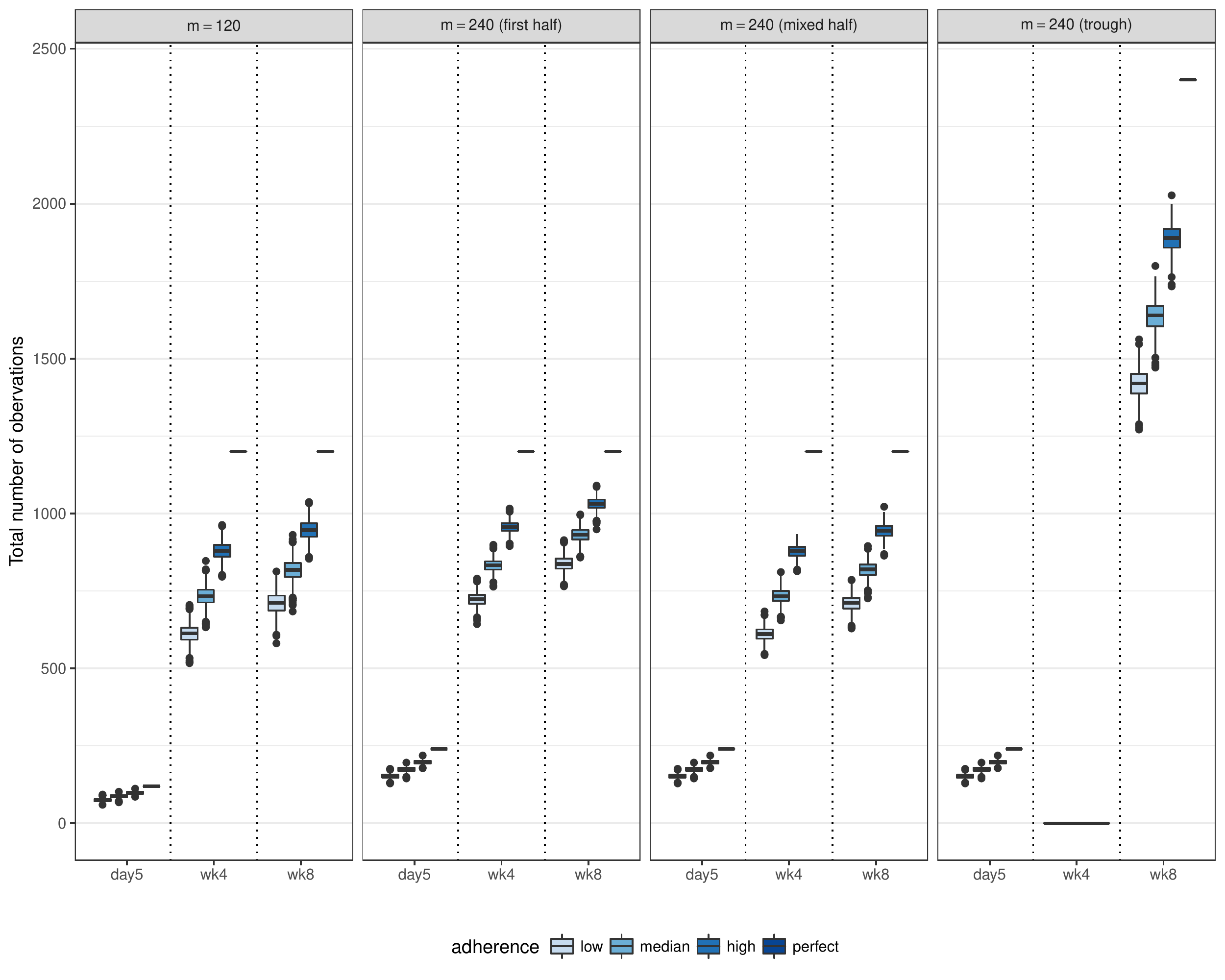}
\end{center}

\newpage
\textbf {Figure 8: Relative bias of each fixed-effect PK parameter estimate under coarsened schedule designs with m = 240 compared to the complete schedule design with m = 120.} $\beta_{CL}$, $\beta_{V_c}$, $\beta_{Q}$, and $\beta_{V_p}$ indicate the fixed effect for CL, $V_c$, Q, and $V_p$, respectively. $\beta_{BW.CL}$ and $\beta_{BW.V_c}$ indicate the fixed effect of body weight influence on CL and $V_c$, respectively. The `First half' design samples the first 11 time-points (excluding time 0) out of the total 22 complete schedule time-points. The `Mixed half' design samples time-points after every other infusion. The `Trough only' design samples only trough time-points. All 3 coarsened schedule designs always include the 5-day post $2^{nd}$ infusion time-point.\\
\begin{center}
\includegraphics[width=0.9\textwidth]{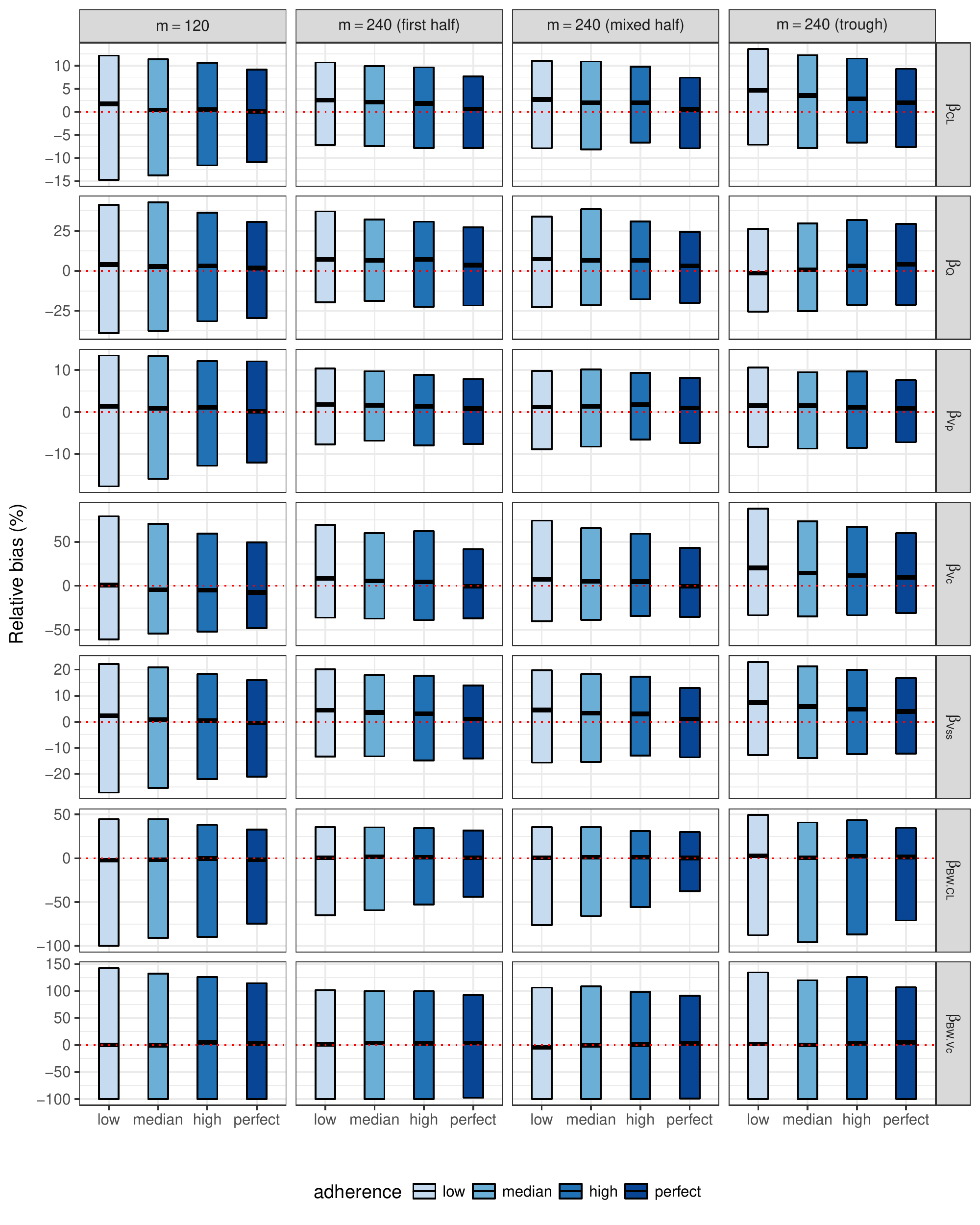}
\end{center}

\newpage
\textbf {Figure 9: Relative root mean squared errors of each fixed-effect PK parameter estimate under coarsened schedule designs with m = 240 compared to the complete schedule design with m = 120.} $\beta_{CL}$, $\beta_{V_c}$, $\beta_{Q}$, and $\beta_{V_p}$ indicate the fixed effect for CL, $V_c$, Q, and $V_p$, respectively. $\beta_{BW.CL}$ and $\beta_{BW.V_c}$ indicate the fixed effect of body weight influence on CL and $V_c$, respectively. The `First half' design samples the first 11 time-points (excluding time 0) out of the total 22 complete schedule time-points. The `Mixed half' design samples time-points after every other infusion. The `Trough only' design samples only trough time-points. All 3 coarsened schedule designs always include the 5-day post $2^{nd}$ infusion time-point.\\
\begin{center}
\includegraphics[width=0.9\textwidth]{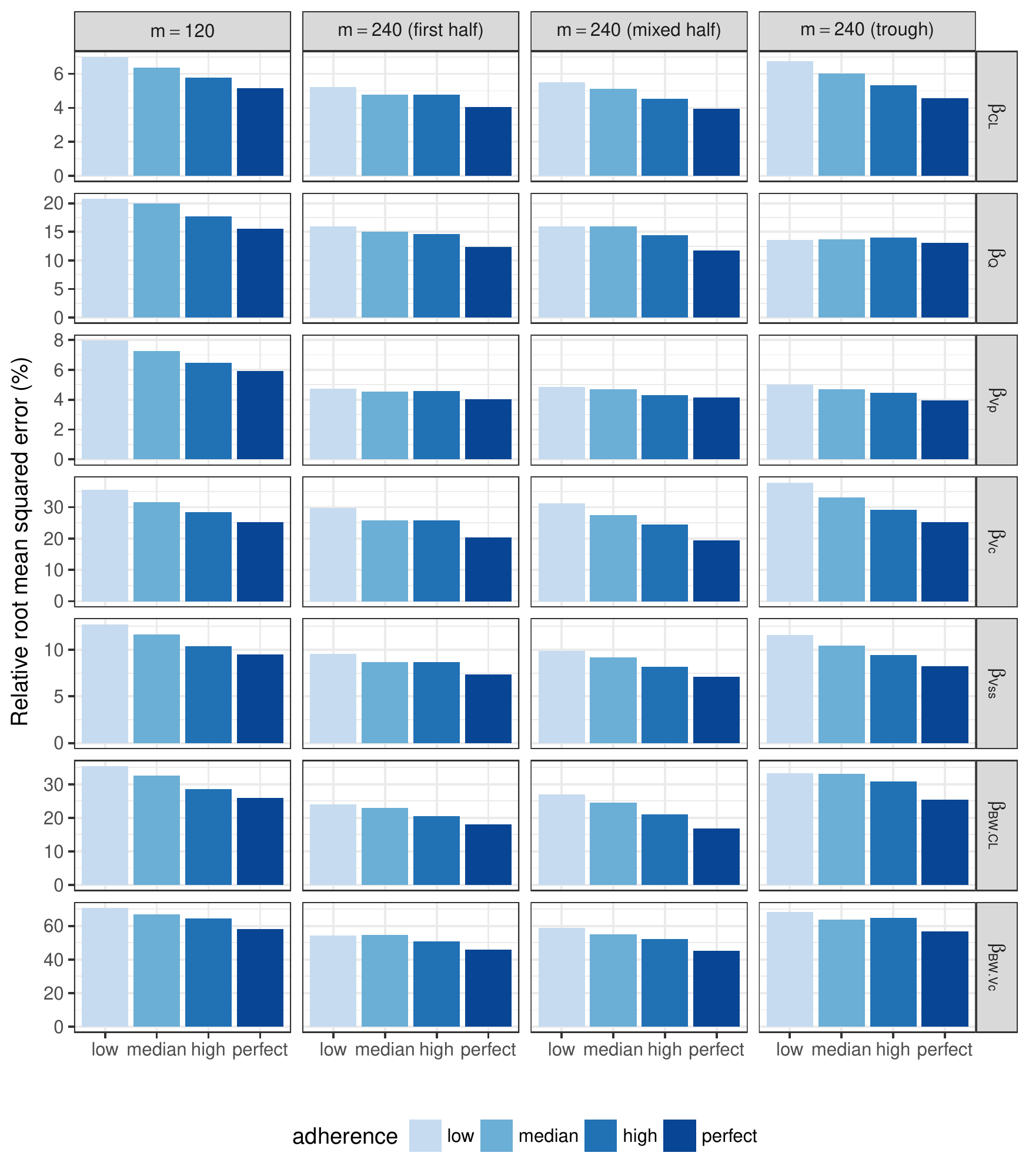}
\end{center}

\newpage 
\section*{Supplemental materials}

\captionsetup{}
\begin{center}
\begin{small}
\begin{longtable}{lclccc}
\caption{\textbf {Fixed effects PK parameter estimation accuracy and precision under complete schedule design scenarios.} $\beta_{CL}$, $\beta_{V_c}$, $\beta_{Q}$, and $\beta_{V_p}$ indicate the fixed effect for CL, $V_c$, Q, and $V_p$, respectively. $\beta_{BW.CL}$ and $\beta_{BW.V_c}$ indicate the fixed effect of body weight influence on CL and $V_c$, respectively. The four study adherence patterns in terms of ($p_1$, $p_2$, $r_1$, $r_2$) are: Perfect = (0\%, 0\%, 0\%, 0\%), High = (5\%, 10\%, 10\%, 10\%), Medium = (10\%, 15\%, 15\%, 15\%), and Low = (15\%, 20\%, 20\%, 20\%), where $p_1=$ probability of an independently missed single infusion, $p_2=$ probability of an independently missed post-infusion visit, $r_1=$ cumulative probability of permanent infusion discontinuation, and $r_2=$ annual drop out rate. RBias $=$ average relative bias, RRMSE $=$ average relative root mean squared error, and CP $=$ coverage probability.}\\

\hline
\hline
Parameter & Sample size & Study adherence & RBias (\%) & RRMSE & CP (\%)\\
\hline
\hline
\endfirsthead
\caption[]{(continued)}\\
\hline
\hline
Parameter & Sample size & Study adherence & Relative bias (\%) & RRMSE & CP (\%)\\
\hline
\hline
\endhead
\hline
 \multirow{16}{*}{$\beta_{CL}$} & \multirow{4}{*}{30} & Low & 11.9	& 26.07	& 95.12\\
  &  & Medium & 7.91	&24.27	&97.54\\
  &  & High & 5.7	&16.84	&97.98\\
  &  & Perfect &4.6	&18.78	&98.14\\
  \cline{2-6}
  & \multirow{4}{*}{60} & Low & 0.43	&10.87	&98.85\\
  &  & Medium &-0.56	&9.7	&99.29\\
  &  & High &-0.45	&9.61	&98.79\\
  &  & Perfect &-1.9	&8.93	&99.4\\
  \cline{2-6}
  & \multirow{4}{*}{120} & Low & 0.88	&6.99	&98\\
  &  & Medium &0.11	&6.37	&99\\
  &  & High &0.18	&5.77	&98.9\\
  &  & Perfect &-0.4	&5.17	&99.4\\
  \cline{2-6}
    & \multirow{4}{*}{240} & Low &1.36	&4.46	&98\\
  &  & Medium &0.69	&4.07	&98.6\\
  &  & High &0.4	&3.99	&99.1\\
  &  & Perfect &-1.23	&3.77	&99.5\\
\hline
 \multirow{16}{*}{$\beta_{V_c}$} & \multirow{4}{*}{30} & Low &35.46	&90.46	&94.7\\
  &  & Medium &17.68	&76.53	&95.47\\
  &  & High &12.83	&68.4	&96.79\\
  &  & Perfect &4.05	&62.86	&96.83\\
  \cline{2-6}
  & \multirow{4}{*}{60} & Low &-1.09	&47.74	&97.49\\
  &  & Medium &-6.91	&42.71	&97.35\\
  &  & High &-5.62	&42.12	&97.37\\
  &  & Perfect &-9.96	&38.26	&97.7\\
  \cline{2-6}
  & \multirow{4}{*}{120} & Low &3.03	&35.48	&98.2\\
  &  & Medium &-1.39	&31.61	&99.1\\
  &  & High &-2.83	&28.33	&98.6\\
  &  & Perfect &-5.51	&25.32	&99.2\\
  \cline{2-6}
    & \multirow{4}{*}{240} & Low & 3.39	&23.28	&99.1\\
  &  & Medium &-1.14	&20.82	&99.4\\
  &  & High & -2.3	&20.38	&98.9\\
  &  & Perfect & -9.65	&19.31	&98.4\\
  \hline
 \multirow{16}{*}{$\beta_{Q}$} & \multirow{4}{*}{30} & Low & 9.63	&53.81	&97.77\\
  &  & Medium &7.13	&50.32	&98.96\\
  &  & High &6.75	&49.24	&99.52\\
  &  & Perfect &2.93	&37.07	&99.45\\
 \cline{2-6}
  & \multirow{4}{*}{60} & Low & -0.06	&31.05	&99.69\\
  &  & Medium &-1.15	&28.55	&99.69\\
  &  & High & -0.43	&27.7	&100\\
  &  & Perfect &-3.51	&24.35	&99.5\\
  \cline{2-6}
  & \multirow{4}{*}{120} & Low & 3.65	&20.78	&99.5\\
  &  & Medium & 2.76	&20.01	&99.5\\
  &  & High & 2.92	&17.74	&99.5\\
  &  & Perfect & 1.37	&15.5	&99.8\\
  \cline{2-6}
    & \multirow{4}{*}{240} & Low & 5.53	&14.18	&99.4\\
  &  & Medium & 4.13	&13.58	&99\\
  &  & High & 3.62	&12.69	&98.9\\
  &  & Perfect &-0.83	&10.86	&99.4\\
  \hline
 \multirow{16}{*}{$\beta_{V_p}$} & \multirow{4}{*}{30} & Low &3.86	&33.95	&99.58\\
  &  & Medium &2.19	&34.38	&99.35\\
  &  & High &0.6	&22.19	&99.64\\
  &  & Perfect &1.08	&25.84	&99.78\\
  \cline{2-6}
  & \multirow{4}{*}{60} & Low &-1.44	&13.76	&99.79\\
  &  & Medium & -1.57	&11.94	&99.49\\
  &  & High &-1.44	&11.95	&99.9\\
  &  & Perfect &-2.14	&10.7	&99.9\\
  \cline{2-6}
  & \multirow{4}{*}{120} & Low &0.64	&7.95	&99.3\\
  &  & Medium &0.43	&7.27	&99\\
  &  & High &0.81	&6.46	&98.8\\
  &  & Perfect &0.27	&5.92	&99\\
  \cline{2-6}
    & \multirow{4}{*}{240} & Low &1.92	&4.77	&97.9\\
  &  & Medium &1.64	&4.49	&98.2\\
  &  & High &1.26	&4.45	&98.3\\
  &  & Perfect &0.13	&3.95	&99\\
  \hline
 \multirow{16}{*}{$\beta_{BW.CL}$} & \multirow{4}{*}{30} & Low &-12.15	&68.68	&97.21\\
  &  & Medium &-17.1	&67.58	&96.89\\
  &  & High &-17.78	&63.76	&98.1\\
  &  & Perfect &-13.88	&60.21	&97.49\\
  \cline{2-6}
  & \multirow{4}{*}{60} & Low &-14.27	&50.84	&89.44\\
  &  & Medium &-17.63	&50.81	&86.85\\
  &  & High &-14.88	&46.7	&86.94\\
  &  & Perfect &-14.61	&43.61	&87.88\\
  \cline{2-6}
  & \multirow{4}{*}{120} & Low &-9.1	&35.43	&86.17\\
  &  & Medium &-7.43	&32.67	&87.2\\
  &  & High &-4.72	&28.63	&90.4\\
  &  & Perfect &-5.11	&25.98	&92.2\\
  \cline{2-6}
    & \multirow{4}{*}{240} & Low &-2.98	&20.97	&92.2\\
  &  & Medium &-3.4	&21.33	&90.4\\
  &  & High &-1.38	&16.18	&95.7\\
  &  & Perfect &-2.24	&13.88	&96.2\\
  \hline
 \multirow{16}{*}{$\beta_{BW.V_c}$} & \multirow{4}{*}{30} & Low &83.58	&175.16	&96.79\\
    &  & Medium &99.46	&405.73	&97.67\\
  &  & High &70.13	&195.68	&97.62\\
  &  & Perfect &69.14	&152.71	&98.14\\
  \cline{2-6}
  & \multirow{4}{*}{60} & Low &28.14	&97.78	&97.8\\
  &  & Medium &22.03	&89.87	&98.27\\
  &  & High &20.75	&90.84	&98.18\\
  &  & Perfect &12.48	&80.15	&97.7\\
  \cline{2-6}
  & \multirow{4}{*}{120} & Low &0.83	&70.7	&98\\
  &  & Medium &-0.23	&66.91	&97.3\\
  &  & High &5.68	&64.54	&96.5\\
  &  & Perfect &1.52	&58.06	&97.4\\
  \cline{2-6}
    & \multirow{4}{*}{240} & Low &-1.85	&51.33	&96.7\\
  &  & Medium &0.5	&50.1	&96.7\\
  &  & High &2.36	&46.26	&95.5\\
  &  & Perfect &1.55	&40.76	&95.6\\
\hline
\hline

\end{longtable}
\end{small}
\end{center}

\begin{center}
\begin{small}
\begin{longtable}{lclccc}
\caption{\textbf {Fixed effects PK parameter estimation accuracy and precision under coarsened schedule design scenarios with n=240.} $\beta_{CL}$, $\beta_{V_c}$, $\beta_{Q}$, and $\beta_{V_p}$ indicate the fixed effect for CL, $V_c$, Q, and $V_p$, respectively. $\beta_{BW.CL}$ and $\beta_{BW.V_c}$ indicate the fixed effect of body weight influence on CL and $V_c$, respectively. The four study adherence patterns in terms of ($p_1$, $p_2$, $r_1$, $r_2$) are: Perfect = (0\%, 0\%, 0\%, 0\%), High = (5\%, 10\%, 10\%, 10\%), Medium = (10\%, 15\%, 15\%, 15\%), and Low = (15\%, 20\%, 20\%, 20\%), where $p_1=$ probability of an independently missed single infusion, $p_2=$ probability of an independently missed post-infusion visit, $r_1=$ cumulative probability of permanent infusion discontinuation, and $r_2=$ annual drop out rate. RBias $=$ average relative bias, RRMSE $=$ average relative root mean squared error, and CP $=$ coverage probability.}\\
\hline
\hline
Parameter & Sampling design & Study adherence & RBias (\%) & RRMSE & CP (\%)\\
\hline
\hline
\endfirsthead
\caption[]{(continued)}\\
\hline
\hline
Parameter & Sampling design & Study adherence & Relative bias (\%) & RRMSE & CP (\%)\\
\hline
\hline
\endhead
\hline
 \multirow{16}{*}{$\beta_{CL}$} & \multirow{4}{*}{m=240 First half} & Low & 2.4	&5.24	&95.5\\
  &  & Medium &1.83	&4.78	&96.9\\
  &  & High &1.6	&4.76	&96.8\\
  &  & Perfect &0.35	&4.05	&98.3\\ 
  \cline{2-6}
  & \multirow{4}{*}{m=240 Mixed half} & Low & 2.23	&5.49	&96.5\\
  &  & Medium &1.8	&5.13	&96.3\\
  &  & High &1.85	&4.54	&97.3\\
  &  & Perfect &0.43	&3.93	&98.9\\
  \cline{2-6}
  & \multirow{4}{*}{m=240 Trough only} & Low & 4.25	&6.74	&92.4\\
  &  & Medium &3.28	&6.02	&93.7\\
  &  & High &2.75	&5.34	&95.2\\
  &  & Perfect &1.8	&4.56	&97.2\\
  \cline{2-6}
    & \multirow{4}{*}{m=120 Full} & Low & 0.88	&6.99	&98\\
  &  & Medium &0.11	&6.37	&99\\
  &  & High &0.18	&5.77	&98.9\\
  &  & Perfect &-0.4	&5.17	&99.4\\ 
\hline
 \multirow{16}{*}{$\beta_{V_c}$} & \multirow{4}{*}{m=240 First half} & Low &11.34	&29.68	&98.6\\
  &  & Medium &7.35	&25.86	&98.7\\
  &  & High &6.4	&25.73	&98.2\\
  &  & Perfect &0.72	&20.42	&98.8\\
  \cline{2-6}
  & \multirow{4}{*}{m=240 Mixed half} & Low &11.51	&31.2	&98.6\\
  &  & Medium &7.51	&27.46	&98.5\\
  &  & High &7.21	&24.45	&99.1\\
  &  & Perfect &0.71	&19.47	&98.9\\
  \cline{2-6}
  & \multirow{4}{*}{m=240 Trough only} & Low &21.97	&37.87	&97.4\\
  &  & Medium &16.57	&33.08	&97.8\\
  &  & High &13.65	&29.21	&98.4\\
  &  & Perfect &10.85	&25.2	&99\\
  \cline{2-6}
    & \multirow{4}{*}{m=120 Full} & Low &3.03	&35.48	&98.2\\
  &  & Medium &-1.39	&31.61	&99.1\\
  &  & High &-2.83	&28.33	&98.6\\
  &  & Perfect &-5.51	&25.32	&99.2\\
  \hline
 \multirow{16}{*}{$\beta_{Q}$} & \multirow{4}{*}{m=240 First half} & Low & 7.71	&15.94	&98.8\\
  &  & Medium &6.69	&15.05	&98.5\\
  &  & High &6.32	&14.64	&97\\
  &  & Perfect &3.21	&12.31	&97.9\\
 \cline{2-6}
  & \multirow{4}{*}{m=240 Mixed half} & Low & 6.79	&15.98	&99.4\\
  &  & Medium &6.47	&15.99	&98\\
  &  & High & 7.05	&14.4	&98.1\\
  &  & Perfect &3.4	&11.71	&98.4\\
  \cline{2-6}
  & \multirow{4}{*}{m=240 Trough only} & Low & -1.04	&13.55	&98.3\\
  &  & Medium & 0.99	&13.71	&98.3\\
  &  & High & 3.52	&13.98	&99\\
  &  & Perfect & 4.12	&13.12	&98.6\\
  \cline{2-6}
    & \multirow{4}{*}{m=120 Full} & Low & 3.65	&20.78	&99.5\\
  &  & Medium & 2.76	&20.01	&99.5\\
  &  & High & 2.92	&17.74	&99.5\\
  &  & Perfect & 1.37	&15.5	&99.8\\
  \hline
 \multirow{16}{*}{$\beta_{V_p}$} & \multirow{4}{*}{m=240 First half} & Low &1.66	&4.75	&97.5\\
  &  & Medium &1.55	&4.54	&97.4\\
  &  & High &1.17	&4.56	&97.7\\
  &  & Perfect &0.68	&4.03	&98\\
  \cline{2-6}
  & \multirow{4}{*}{m=240 Mixed half} & Low &1.1	&4.86	&98.4\\
  &  & Medium & 1.24	&4.71	&97.4\\
  &  & High &1.47	&4.3	&97.8\\
  &  & Perfect &0.78	&4.13	&98.2\\
  \cline{2-6}
  & \multirow{4}{*}{m=240 Trough only} & Low &1.37	&5.01	&98\\
  &  & Medium &1.21	&4.69	&97.9\\
  &  & High &1.1	&4.48	&98.5\\
  &  & Perfect &0.77	&3.94	&98.8\\
  \cline{2-6}
    & \multirow{4}{*}{m=120 Full} & Low &0.64	&7.95	&99.3\\
  &  & Medium &0.43	&7.27	&99\\
  &  & High &0.81	&6.46	&98.8\\
  &  & Perfect &0.27	&5.92	&99\\
  \hline
 \multirow{16}{*}{$\beta_{BW.CL}$} & \multirow{4}{*}{m=240 First half} & Low &-2.11	&23.94	&90.3\\
  &  & Medium &-1.6	&22.91	&89.7\\
  &  & High &-0.68	&20.43	&91.4\\
  &  & Perfect &-1.14	&18.12	&93\\
  \cline{2-6}
  & \multirow{4}{*}{m=240 Mixed half} & Low &-4.17	&27	&87.7\\
  &  & Medium &-2.49	&24.61	&89.6\\
  &  & High &-1.09	&21.12	&92.2\\
  &  & Perfect &-0.63	&16.9	&94.3\\
  \cline{2-6}
  & \multirow{4}{*}{m=240 Trough only} & Low &-3.73	&33.35	&85.6\\
  &  & Medium &-5.87	&33.07	&86\\
  &  & High &-3.31	&30.86	&88.3\\
  &  & Perfect &-2.91	&25.49	&89.8\\
  \cline{2-6}
    & \multirow{4}{*}{m=120 Full} & Low &-9.1	&35.43	&86.17\\
  &  & Medium &-7.43	&32.67	&87.2\\
  &  & High &-4.72	&28.63	&90.4\\
  &  & Perfect &-5.11	&25.98	&92.2\\
  \hline
 \multirow{16}{*}{$\beta_{BW.V_c}$} & \multirow{4}{*}{m=240 First} & Low &-2	&54.42	&96.7\\
    &  & Medium &-1.1	&54.68	&96.2\\
  &  & High &0.84	&50.76	&95.6\\
  &  & Perfect &2.45	&45.73	&95.2\\
  \cline{2-6}
  & \multirow{4}{*}{m=240 Mixed half} & Low &-5.88	&58.98	&96.7\\
  &  & Medium &-2.13	&55.02	&95.8\\
  &  & High &-1.15	&52.32	&95.7\\
  &  & Perfect &2.98	&45.23	&94.3\\
  \cline{2-6}
  & \multirow{4}{*}{m=240 Trough only} & Low &2.14	&68.17	&94.8\\
  &  & Medium &-2.06	&63.8	&95.1\\
  &  & High &2.49	&64.63	&94.4\\
  &  & Perfect &1.45	&56.91	&96.3\\
  \cline{2-6}
    & \multirow{4}{*}{m=120 Full} & Low &0.83	&70.7	&98\\
  &  & Medium &-0.23	&66.91	&97.3\\
  &  & High &5.68	&64.54	&96.5\\
  &  & Perfect &1.52	&58.06	&97.4\\
\hline
\hline
\end{longtable}
\end{small}
\end{center}

\newpage
\textbf {Figure S1: Relative bias of each random-effect PK parameter estimate under complete schedule designs with different sample sizes (m = 60, 120, and 240).} $\mbox{var}(b_{CL})$, $\mbox{var}(b_{Q})$, and $\mbox{var}(b_{V_p})$ are the variances of the random effects for CL, Q, and $V_p$, respectively. $\mbox{cov}(b_{CL}, b_{Q})$, $\mbox{cov}(b_{CL}, b_{V_p})$, and $\mbox{cov}(b_{CL}, b_{V_c})$ are the covariances between the respective random effects. $\sigma^2_1$ and $\sigma^2_2$ are the proportional and additive error variances, respectively.\\
\begin{center}
\includegraphics[width=0.9\textwidth]{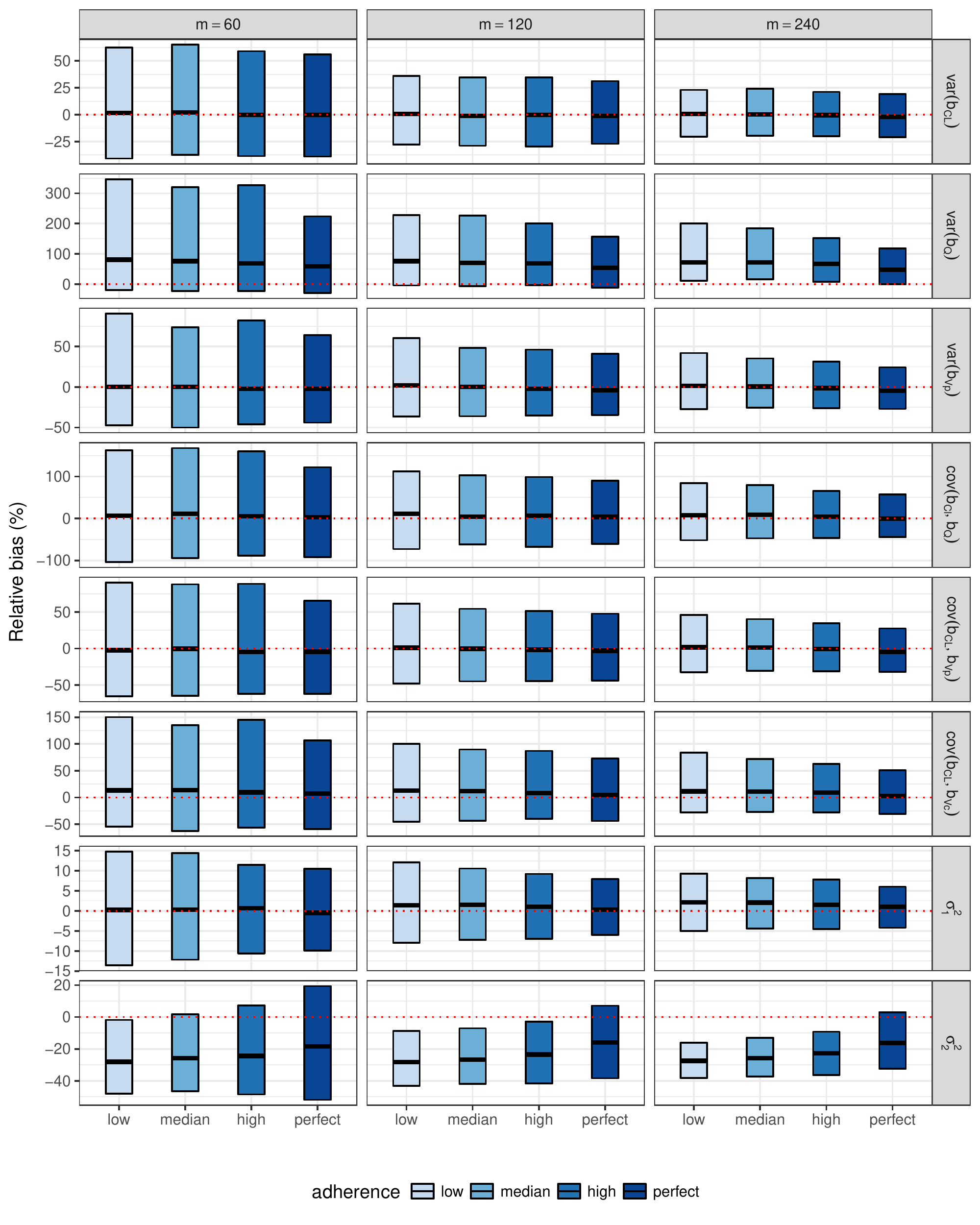}
\end{center}

\newpage
\textbf {Figure S2: Relative root mean squared errors of each random-effect PK parameter estimate under complete schedule designs with different sample sizes (m = 60, 120, and 240).} $\mbox{var}(b_{CL})$, $\mbox{var}(b_{Q})$, and $\mbox{var}(b_{V_p})$ are the variances of the random effects for CL, Q, and $V_p$, respectively. $\mbox{cov}(b_{CL}, b_{Q})$, $\mbox{cov}(b_{CL}, b_{V_p})$, and $\mbox{cov}(b_{CL}, b_{V_c})$ are the covariances between the respective random effects. $\sigma^2_1$ and $\sigma^2_2$ are the proportional and additive error variances, respectively.\\
\begin{center}
\includegraphics[width=0.9\textwidth]{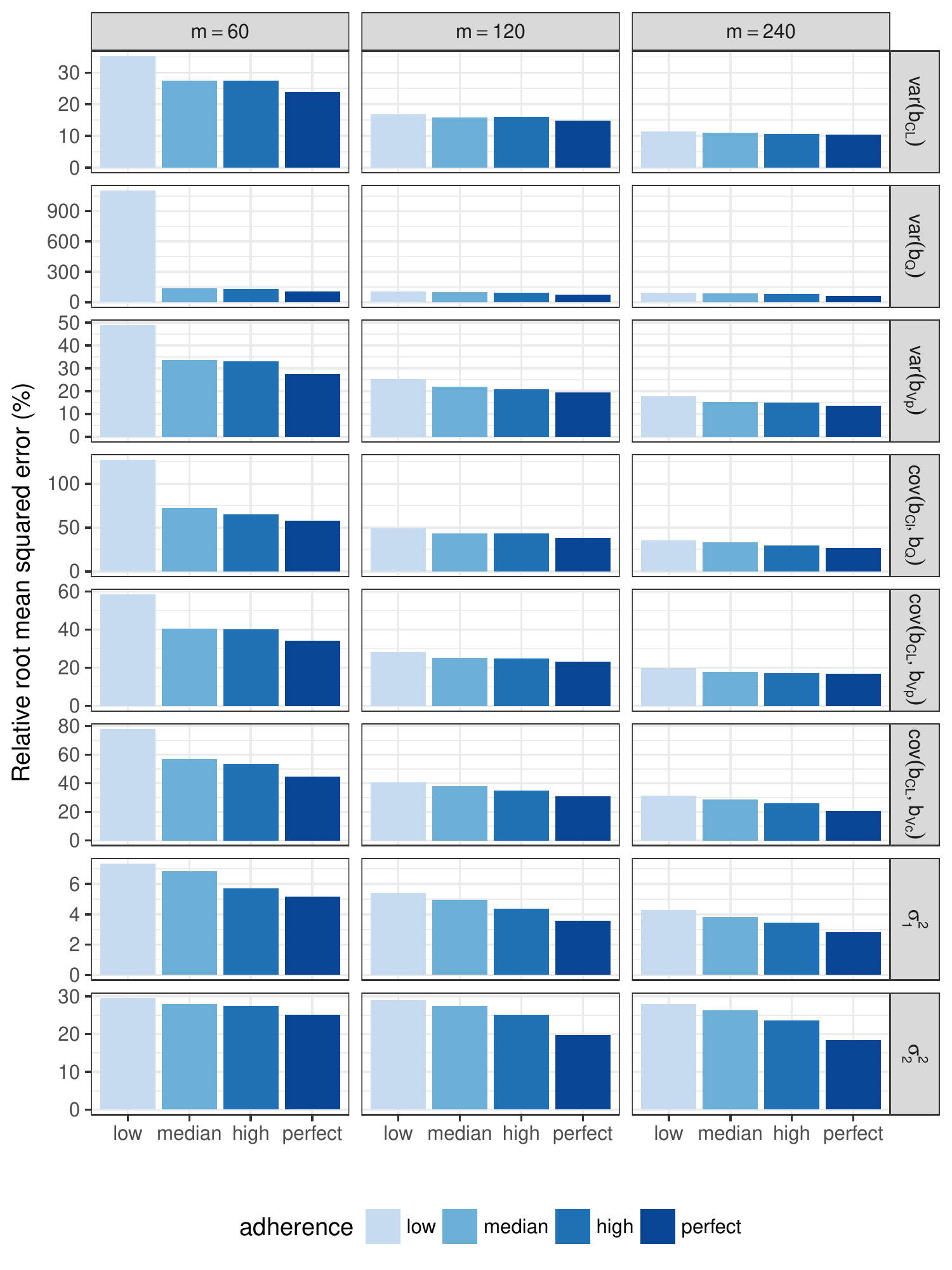}
\end{center}

\newpage
\textbf {Figure S3: Shrinkage estimates under complete schedule design scenarios.} $b_{CL}$, $b_{Q}$, and $b_{V_p}$ are the random effects for CL, Q, and $V_p$, respectively.\\
\begin{center}
\includegraphics[width=1.0\textwidth]{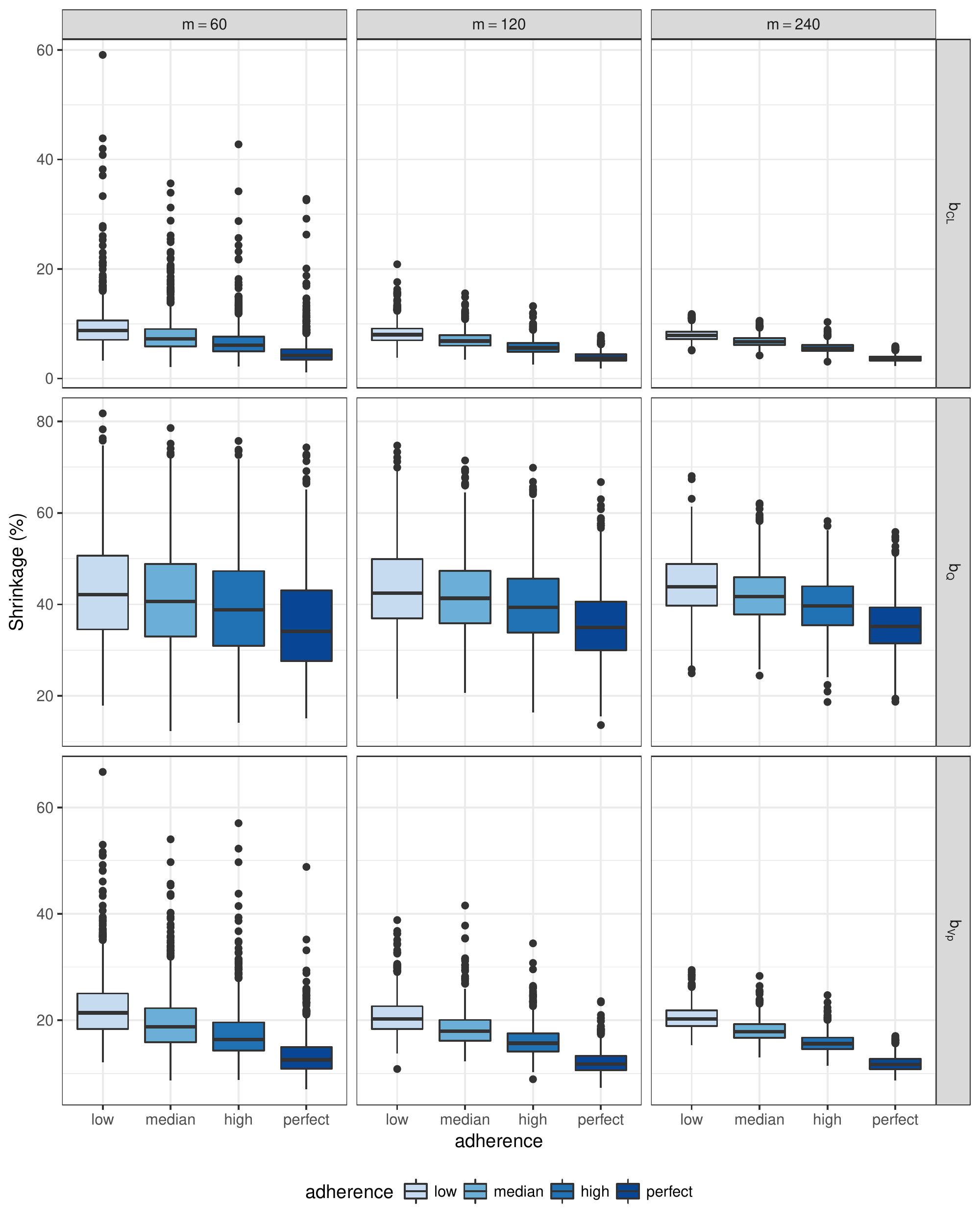}
\end{center}

\newpage
\textbf {Figure S4: Relative bias of each random-effect PK parameter estimate under coarsened schedule designs with m = 240 compared to the complete schedule design with m = 120.} $\mbox{var}(b_{CL})$, $\mbox{var}(b_{Q})$, and $\mbox{var}(b_{V_p})$ are the variances of the random effects for CL, Q, and $V_p$, respectively. $\mbox{cov}(b_{CL}, b_{Q})$, $\mbox{cov}(b_{CL}, b_{V_p})$, and $\mbox{cov}(b_{CL}, b_{V_c})$ are the covariances between the respective random effects. $\sigma^2_1$ and $\sigma^2_2$ are the proportional and additive error variances, respectively.\\
\begin{center}
\includegraphics[width=0.9\textwidth]{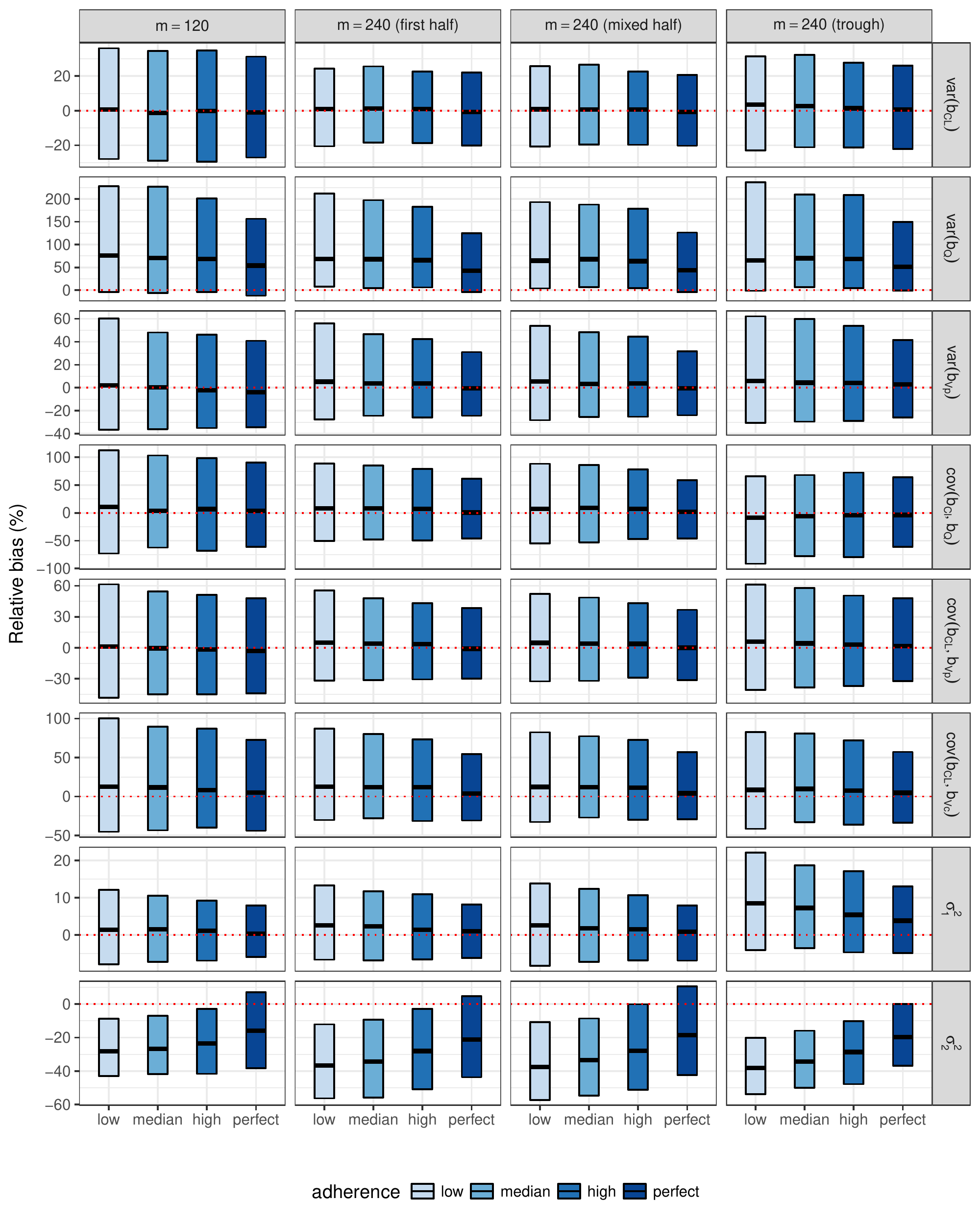}
\end{center}

\newpage
\textbf {Figure S5: Relative root mean squared errors of each random-effect PK parameter estimate under coarsened schedule designs with m = 240 compared to the complete schedule design with m = 120.} $\mbox{var}(b_{CL})$, $\mbox{var}(b_{Q})$, and $\mbox{var}(b_{V_p})$ are the variances of the random effects for CL, Q, and $V_p$, respectively. $\mbox{cov}(b_{CL}, b_{Q})$, $\mbox{cov}(b_{CL}, b_{V_p})$, and $\mbox{cov}(b_{CL}, b_{V_c})$ are the covariances between the respective random effects. $\sigma^2_1$ and $\sigma^2_2$ are the proportional and additive error variances, respectively.\\
\begin{center}
\includegraphics[width=1.0\textwidth]{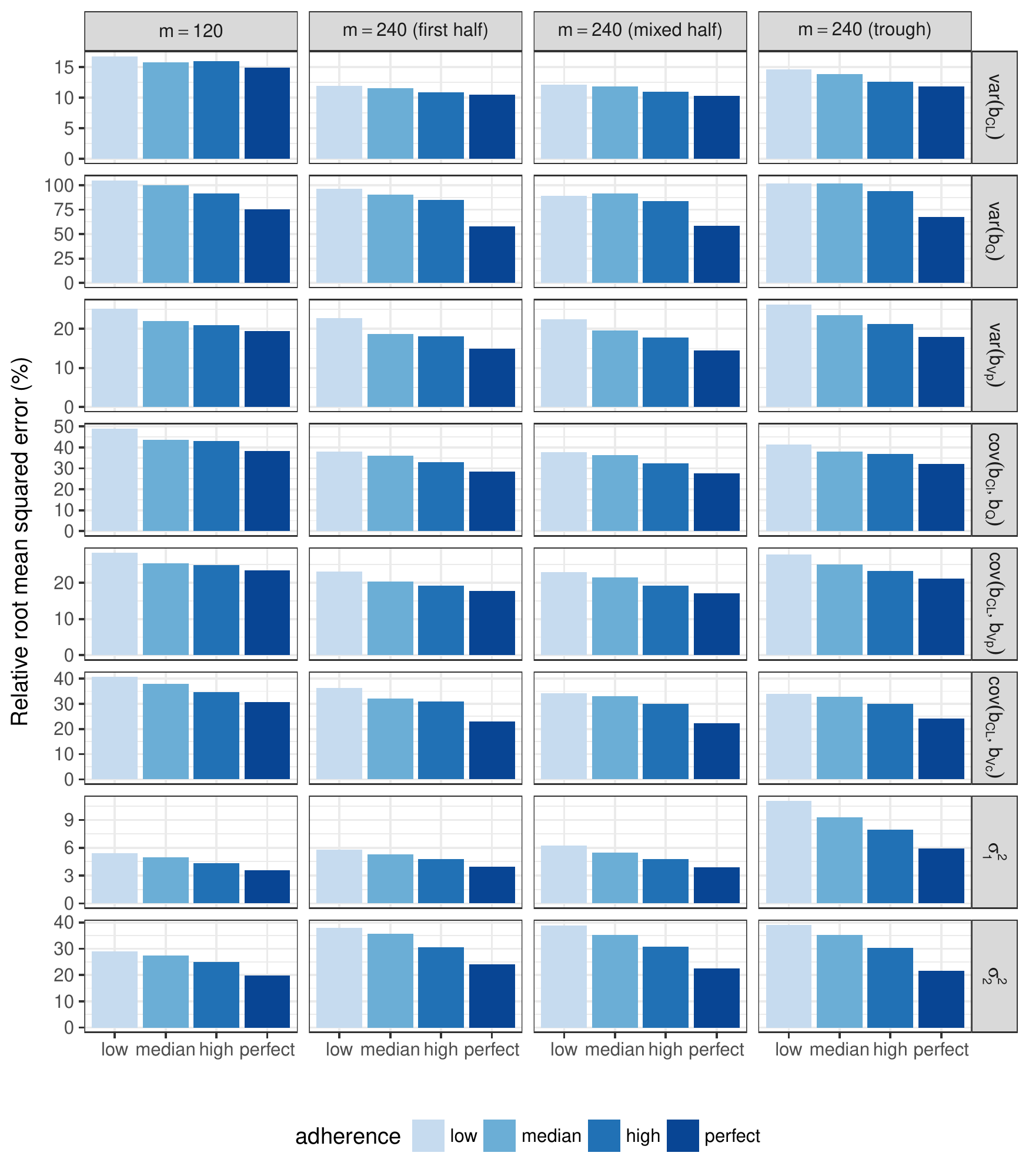}
\end{center}

\newpage
\textbf {Figure S6: Shrinkage estimates under coarsened schedule designs with m = 240 compared to the complete schedule design with m = 120.} The `First half' design samples the first 11 time-points (excluding time 0) out of the total 22 complete schedule time-points. The `Mixed half' design samples time-points after every other infusion. The `Trough only' design samples only trough time-points. All 3 coarsened schedule designs always include the 5-day post $2^{nd}$ infusion time-point. $b_{CL}$, $b_{Q}$, and $b_{V_p}$ are the random effects for CL, Q, and $V_p$, respectively.\\
\begin{center}
\includegraphics[width=1.0\textwidth]{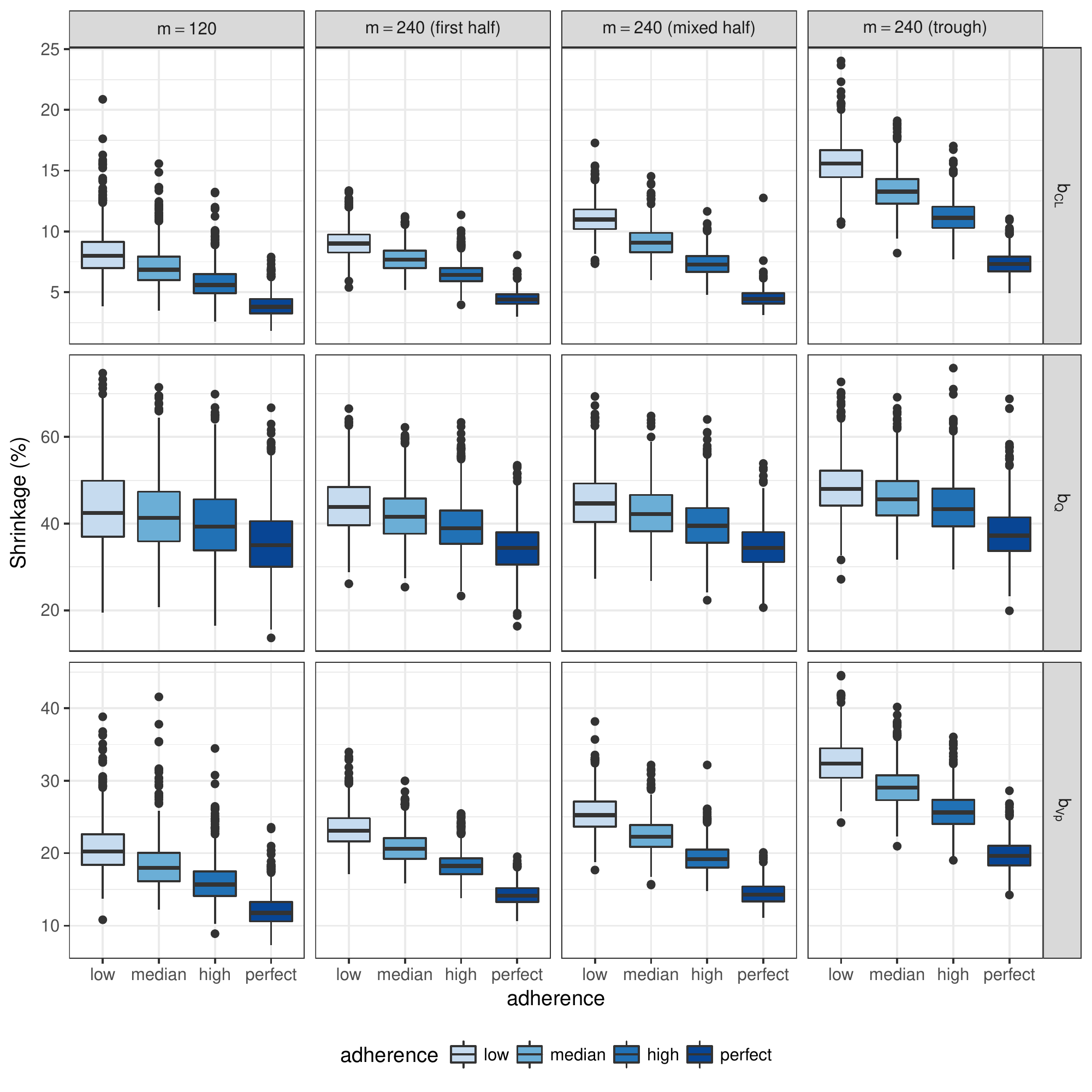}
\end{center}

\begin{center}


\section*{Supplementary material}

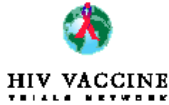 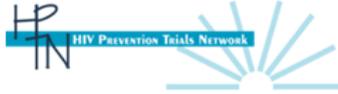

# HVTN 704/HPTN 085

Study Specific Procedures

# Visit Scheduling and Coding

February 26, 2016

To ensure completeness and accuracy of data submitted to the statistical and data management center (SDMC), proper visit scheduling and coding is essential. This section describes the processes involved in visit scheduling and coding of study visits on case report forms (CRFs).

1. **Visit Schedules**

   There will be four separate visit schedule tracks in the HVTN 704/HPTN 085 study:

   - Schedule 1: HIV-uninfected participants

   - Schedule 2: HIV-1-infected participants

   - Schedule 3: Participants discovered to have been HIV-1-infected at enrollment or who become HIV-2-infected

   - Schedule 4: Participants who discontinue infusions for reasons other than HIV infection

   Infusion visits will only occur within Schedule 1.

2. **Study Visit Timing**

   All visits should take place within the "target visit windows". Infusion visits and some Schedule 2, 3, and 4 visits have an expanded "allowable visit window" that *may* be utilized <u>only if the visit cannot be scheduled within the target visit window</u>. The visit windows are used by DataFax to query for an "overdue" visit and in the Retention Reports to determine whether a visit has been conducted "early" or "late."

2.1. Visit Windows

The visit windows for each HVTN 704/HPTN 085 schedule are listed in Tables 1-4.

Sites should verify visit windows prior to scheduling participant visits and again prior to conducting the visit. This should be done by using the HVTN 704/HPTN 085 Visit Calculator spreadsheet, posted on the HVTN 704/HPTN 085 home page: https://members.hvtn.org/protocols/hvtn704-hptn085/SitePages/Home.aspx or, if unavailable, then by using another reliable mechanism such as a Scheduling Wheel-Chart. As visit windows for HIV-uninfected participants are dynamic and based on

the prior infusion date (see Section 2.1.1 below), the most current participant visit schedule should be used (not a static printout).

If more than one infusion visit for a single participant occurs within one 49 day period this is considered a protocol event. Submit a Protocol Event and Critical Event form (PECEF) to the Clinical Trials Manager (CTM)/Clinical Research Manager (CRM) (vtn704.ptn085.ctm.crm@hvtn.org) and complete the Protocol Event Log CRF (refer to the SSP section on Protocol Events and Critical Events for more information).

2.1.1. Visit Windows for Participants in the HIV-Uninfected Schedule (Schedule 1)

An idealized visit schedule is depicted in Appendix J of the HVTN 704/HPTN 085 protocol Version 1.0; however, visit windows are dynamic and must always be determined according to the date the prior infusion visit occurred (as such, it is expected that Appendix J may not be an appropriate guide for visit windows for most of a participant's study visits).

Infusion Visits

**All infusion visits must be at least 49 days apart.** Target days, target windows, and allowable windows of an infusion are based upon the date of the prior infusion visit. If an infusion visit is missed, the subsequent infusion should be scheduled within a visit window that is based on the target day of the prior (missed) infusion. Sites should attempt to schedule infusion visits within the target visit window but have the option of utilizing the upper allowable window if needed. There is no "lower allowable window" for an infusion visit: the target window for any infusion visit encompasses the earliest possible dates at which the infusion may be conducted; the "upper allowable window" for any infusion visit encompasses the latest possible dates at which the infusion may be conducted.

Post-Infusion Visits

All post-infusion visits are scheduled according to the date of the prior infusion visit. When an infusion visit is missed, the 5-day post-infusion visit (if applicable) and 4-week post-infusion visit are also considered missed (see Section 2.4 below). If visit 21 (infusion #10) is missed, visits 23-26 should be scheduled based on the date of the target day of the missed visit 21. Subsequent infusion visit windows are not impacted if a post-infusion visit is missed.

Scenarios

Refer to HVTN 704/HPTN 085 Visit Scheduling and Coding SSP Appendix, Visit Window Scenarios and Schematics for an illustration of how the dynamic visit windows can be modified for different scenarios during the trial.

Table 1: Visit Windows for HIV-uninfected Participants (Schedule 1)

| Visit Number(s) | Visit Type | Lower Target/ Allowable Window | Target Day | Upper Target Window | Upper Allowable Window |
|---|---|---|---|---|---|
| 01.0 | Screening | -56[1] | | - | - |
| 02.0 | Enrollment/Infusion #1[1] | - | 0 | - | - |
| 04.0 07.0 09.0 11.0 13.0 15.0 17.0 19.0 21.0 | Infusions #2-10[2] | -7 | prior infusion visit date + 56 days | +7 | +48 |
| 05.0 | 5-day post-infusion #2[3] | -2 | Visit 4 date + 5 days | +2 | - |
| 03.0 06.0 08.0 10.0 12.0 14.0 16.0 18.0 20.0 22.0 | 4-week post-infusion #2-10[3] | -7 | prior infusion visit date + 28 days | +7 | - |
| 23.0 | 8-week post-infusion #10[3] | -7 | Visit 21 date + 56 days | +7 | - |
| 24.0 | 12-week post- infusion #10[3] | -7 | Visit 21 date + 84 days | +7 | - |
| 25.0 | 16-week post –infusion #10[3] | -7 | Visit 21 date + 112 days | +7 | - |
| 26.0 | Final Visit (20-week post-infusion #10)[3] | -7 | Visit 21 date + 140 days | +7 | - |

[1] Screening HIV testing must be performed within 30 days of Enrollment/Infusion #1.
[2] Infusions must occur at least 49 days apart and are scheduled according to the date of the prior infusion visit.
[3] Post-infusion visits are scheduled according to the date of the prior infusion visit.

2.1.2. Visit Windows for Participants in the HIV-infected Schedules (Schedule 2 and Schedule 3)

All target days are relative to the date of HIV infection diagnosis. The date of HIV diagnosis is defined as the specimen collection date of the first sample that leads to a positive result by the HIV diagnostic algorithm. Refer to the HIV Testing and Counseling SSP section for further guidance on the HIV testing algorithm, and the Participant Follow-up SSP section for further guidance on transitioning to Schedule 2 or 3.

Table 2: Visit Windows for HIV-1-infected Participants (Schedule 2)

| Visit Number | Visit Type | Lower Allowable Window | Lower Target Window | Target Day (days after diagnosis) | Upper Target Window | Upper Allowable Window |
|---|---|---|---|---|---|---|
| #.x (Interim visit) | Confirmatory draw (HIV diagnostics) (2 weeks after diagnosis) | - | -12 | 14 | +7 | - |
| 31.0* | 4-weeks post diagnosis* | - | -4* | 28* | +4* | - |
| 32.0 | 6-weeks post diagnosis | - | -4 | 42 | +4 | - |
| 33.0 | 8-weeks post diagnosis | - | -4 | 56 | +4 | - |
| 34.0 | 12-weeks post diagnosis | -14 | -7 | 84 | +7 | +14 |
| 35.0 | 24-weeks post diagnosis (Final Visit) | -28 | -14 | 168 | +14 | +28 |

Note: All visits are relative to the HIV infection diagnosis date (not shown on this table).
* Note that confirmatory HIV test results may or may not be available by the opening of the 4-week post diagnosis visit window. Sites should verify availability of confirmatory HIV test results with the HVTN Laboratory Program if scheduling this visit prior to the target day. Questions should be directed to: VTN704.PTN085.askHIV@hvtn.org.

Table 3: Visit Windows for Participants discovered to have been HIV-1-infected at Enrollment or who become HIV-2 infected (Schedule 3)

| Visit Number | Visit Type | Lower Allowable Window | Lower Target Window | Target Day (days after diagnosis) | Upper Target Window | Upper Allowable Window |
|---|---|---|---|---|---|---|
| #.x (Interim visit) | Confirmatory draw (HIV diagnostics) (2 weeks after diagnosis) | - | -7 | 14 | +7 | - |
| 47.0** | 4-weeks post diagnosis** | - | -4** | 28** | +14 | - |
| 48.0 | 24-weeks post diagnosis (Final Visit) | -28 | -14 | 168 | +14 | +28 |

Note: All visits are relative to the HIV infection diagnosis date (not shown on this table).
**Sites should verify availability of confirmatory HIV test results with the HVTN Laboratory Program if scheduling this visit prior to the target day. Questions should be directed to: VTN704.PTN085.askHIV@hvtn.org.

2.1.3. Visit Windows for Participants who Discontinue Infusions for Reasons Other Than HIV Infection (Schedule 4)

All target days are relative to the date of Enrollment. Once a participant has been administratively transitioned to Schedule 4 (see Participant Follow-up SSP section for more information), the participant should be scheduled for the next consecutive study visit within Schedule 4 based on the date (i.e. study day) of the previous visit in Schedule 1. For example, if a participant's infusions are discontinued at the Enrollment visit (Day 0), their next study visit in Schedule 4 should be visit 71 (8 week post enrollment visit), scheduled around target day 56. If a participant's infusions are discontinued on day 190, their next study visit in Schedule 4 should be visit 74 (44 weeks post enrollment), scheduled around target day 308. This is because day 190 (date of discontinuation) is within the visit window for visit 73 (32 weeks post enrollment), so the next consecutive study visit is visit 74. Sites are encouraged to use the AMP Visit Calculator Excel spreadsheet tool to determine the point of entry into Schedule 4.

If a participant is transitioned to Schedule 4 due to a clinical or safety concern, additional interim visits (see Section 2.3 below), in addition to the visits reflected in Schedule 4, may be appropriate. The clinical safety staff (Clinical Safety Specialist [CSS]/Regional Medical Liaison [RML]) should be contacted at VTN704.PTN085.CSS.RML@hvtn.org if a participant is transitioned to Schedule 4 due to a clinical or safety concern. They will work with the HVTN 704/HPTN 085 Protocol Safety Review Team (PSRT) and the Clinical Research Site (CRS) to determine the most appropriate follow-up schedule.

Table 4: Visit Windows for Participants Who Discontinue Infusions for Reasons Other Than HIV Infection (Schedule 4)

| Visit Number | Visit Type | Lower Allowable Window | Lower Target Window | Target Day (days after enrollment) | Upper Target Window | Upper Allowable Window |
|---|---|---|---|---|---|---|
| 71.0 | 8 weeks post enrollment | -55 | -14 | 56 | +14 | +42 |
| 72.0 | 20 weeks post enrollment | -41 | -14 | 140 | +14 | +42 |
| 73.0 | 32 weeks post enrollment | -41 | -14 | 224 | +14 | +42 |
| 74.0 | 44 weeks post enrollment | -41 | -14 | 308 | +14 | +42 |
| 75.0 | 56 weeks post enrollment | -41 | -14 | 392 | +14 | +42 |
| 76.0 | 68 weeks post enrollment | -41 | -14 | 476 | +14 | +42 |
| 77.0 | 80 weeks post enrollment | -41 | -14 | 560 | +14 | +42 |
| 78.0 | 92 weeks post enrollment | -41 | -14 | 644 | +14 | +42 |

Note: All visits are relative to enrollment date

## 2.2. Split Visits

When a participant is not able to complete all required visit procedures on the same day, the participant may return and complete the remaining procedures on another day, as long as the procedures are completed within the allowable window for that visit.

CRS staff must make every effort to complete the infusion visit in one calendar day. To avoid issues from split infusion visits, such as repeat blood draws or blood draw volumes exceeding the 56-day blood volume maximum, it is recommended that, prior to collecting blood samples, sites perform all procedures and evaluations necessary to confirm that the participant can proceed with infusion. This includes conducting the abbreviated physical exam and pregnancy testing (if required) and soliciting adverse events (AEs) and other possible contraindications to infusion.

However, when a split visit must occur at an infusion visit (e.g., due to intercurrent illness or other contraindication to receipt of study product) follow the guidance below:

- Pregnancy test, if required, must be repeated on the day of infusion, and the result confirmed negative prior to infusion, (see Specimen Collection SSP, Pregnancy Testing Section 5).

- Blood specimens (immunogenicity, safety labs, and HIV testing) must be collected within 3 days prior to infusion. If blood has already been collected and infusion must be delayed, attempt to bring participant back for infusion within 3 days to avoid re-draw. If participant does not return within 3 days,

> blood will need to be redrawn according to the Specimen Collection SSP sections on Specimen Collection Priorities and Instructions for Short Draws, taking into consideration the 56-day blood draw volume limitation.
>
> - Physical exams must be performed on the day of infusion.

When a split visit occurs, all CRFs completed for the visit are assigned the same visit code, even though some forms or procedures will have different visit dates. For Schedule 1 visits, the subsequent post-infusion visit window and subsequent infusion visit window will be based on the date the infusion was performed.

**2.3.** Interim Visits

In addition to the scheduled protocol-required visits, interim visits may occur after enrollment in the trial. A clinic visit is considered an interim visit when study data are collected on CRFs during a clinic visit that occurs for reasons *other* than to complete regularly scheduled (required) study visit procedures, or, when allowed (see Section 2.4 below), to complete procedures for a regular study visit that could not be conducted during the visit window.

Interim visits may be performed at any time during the study for reasons that are administrative (e.g., a participant has study-related questions for the staff), lab-related (e.g., a participant needs a lab test repeated for confirmation), or clinical (e.g., a participant needs additional clinical follow-up for an AE). Phone or e-mail contact with a participant can also be considered an interim visit if study data are collected on a CRF.

If no CRFs are required (e.g., the participant comes to the clinic to obtain more condoms), this is not considered an interim visit and you should not complete the Interim Visit CRF; however, the contact should be documented by a chart note.

2.3.1. Interim Visit Codes

Interim visit codes are assigned using the following guidelines:

- In the boxes to the left of the decimal point, record the two-digit visit code for the most recent scheduled visit (even if a Missed Visit CRF was completed for that visit).

- Use the guide below to complete the box to the right of the decimal point:

1 = the first interim visit after the most recent scheduled visit,

2 = the second interim visit after the most recent scheduled visit,

3 = the third interim visit after the most recent scheduled visit, and so on.

For example, the first interim visit after the Enrollment Visit would be recorded as follows:

Visit Code: **02.1**

The second interim visit after the participant's third scheduled study visit would be recorded as follows:

Visit Code: **03.2**

**2.4.** Missed Visits

A visit not performed within the specified visit windows (see above Section 2.1) is considered a missed visit and a Missed Visit CRF must be submitted for that visit. The Missed Visit CRF should only be completed and submitted for a visit once the visit window for that visit has closed. Below is additional guidance for specific types of missed visits.

Infusion Visits

If an infusion visit's allowable window closes and the infusion has not been completed, the infusion visit is considered missed. If an infusion visit is missed, the associated post-infusion visit is automatically considered missed as well, and the subsequent infusion visit window should be determined based on the **target day** of the prior (missed) infusion. CRS staff should attempt to bring the participant in to complete the next infusion visit.

Post-Infusion Visits

If the missed visit is a post-infusion visit **for an infusion visit that was completed**, a Missed Visit CRF should be submitted to DataFax. If possible, the CRS staff should attempt to bring the participant in for an interim visit. At the interim visit, all post-infusion visit protocol-required procedures should be conducted.

If the missed visit is a post-infusion visit due to a **missed infusion visit**, a Missed Visit CRF does NOT need to be submitted for the missed post-infusion visit (a missed infusion visit automatically turns off the expectation in DataFax for the post-infusion visit). CRS staff should attempt to bring the participant in to complete the next infusion visit as soon as possible.

Schedule 2 and 3 Visits

If the missed visit is a Schedule 2 or Schedule 3 visit, CRS staff should attempt to bring the participant in for an interim visit. At the interim visit, all protocol-required procedures for the missed visit should be conducted.

Schedule 4 Visits

As Schedule 4 visit windows are adjacent, if a Schedule 4 visit is missed, the CRS staff should attempt to bring the participant in for the next Schedule 4 visit as soon as possible.

3. **Termination and Reactivation**

Participants are terminated from study participation when they have completed the final study visit, died, refused further participation in the study, relocated, CRS staff can no longer contact the participant (lost-to-follow-up), or the investigator has decided they can or should no longer participate. Participants who have been terminated from the study due to lost-to-follow-up, refusal, or relocation and then return to the study site before the close of the visit window of their exit visit may be allowed to rejoin the study, and should be encouraged to do so. Participants are reactivated by writing "unterminate" or "delete" on the previously submitted Termination CRF and retransmitting to DataFax. Please note that a Missed Visit CRF must be completed for each missed visit during the period the participant was not in contact with the site. Refer to Last Scheduled Study Visit section of the SSP for more information.

4. **Version History**

| February 26, 2016 | Original version |
| --- | --- |
|  |  |

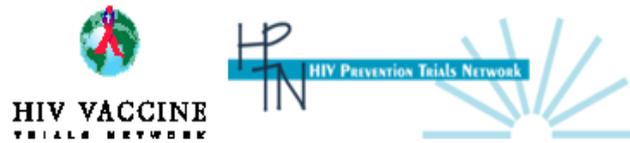

# Appendix:
# HVTN 704/HPTN 085 Visit Scheduling and Coding SSP

*Visit Window Scenarios and Schematics*

# Scenario 1:
# Participant completes visits on target dates in middle of target windows

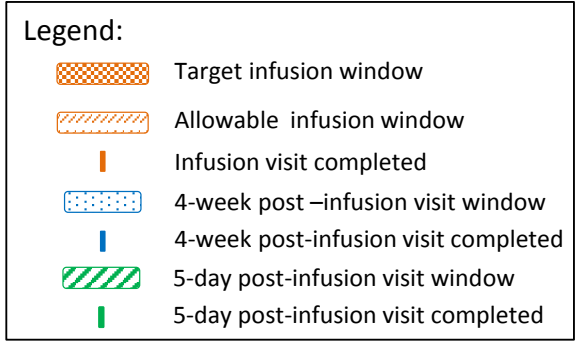

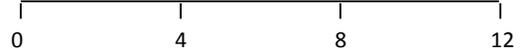

A. Ppt completes 1st infusion

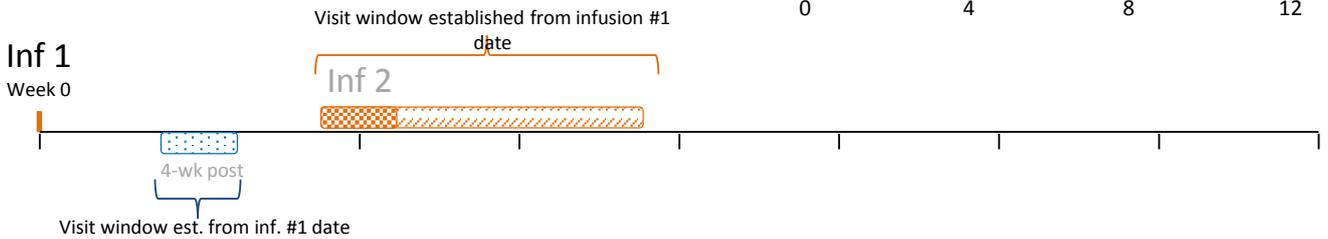

B. Ppt completes 4-week post after 1st infusion

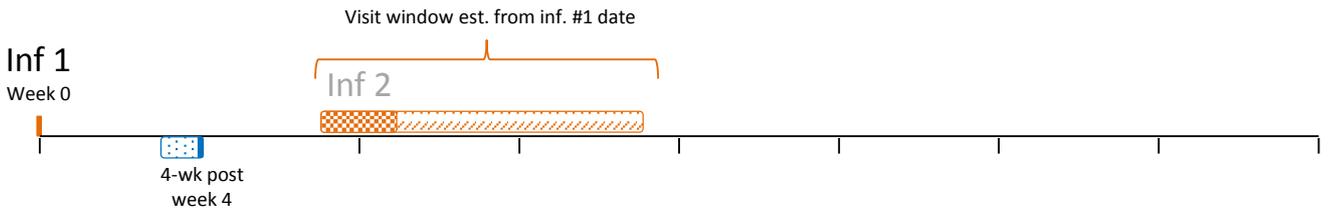

C. Ppt completes 2nd infusion

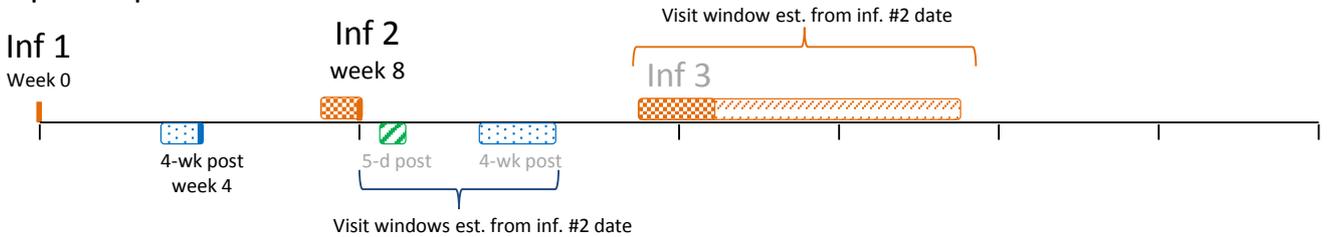

D. Ppt completes 4-week post after 2nd infusion

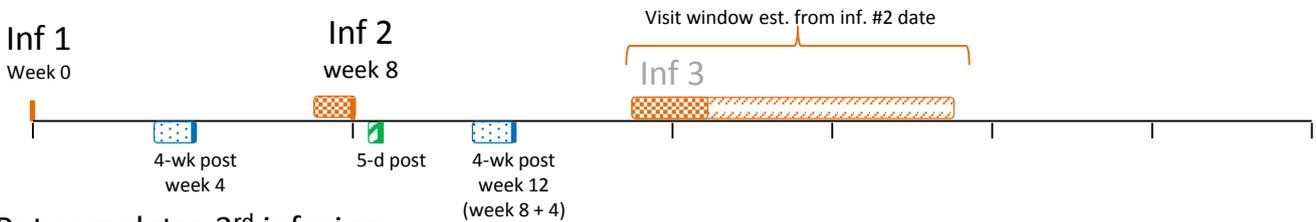

E. Ppt completes 3rd infusion

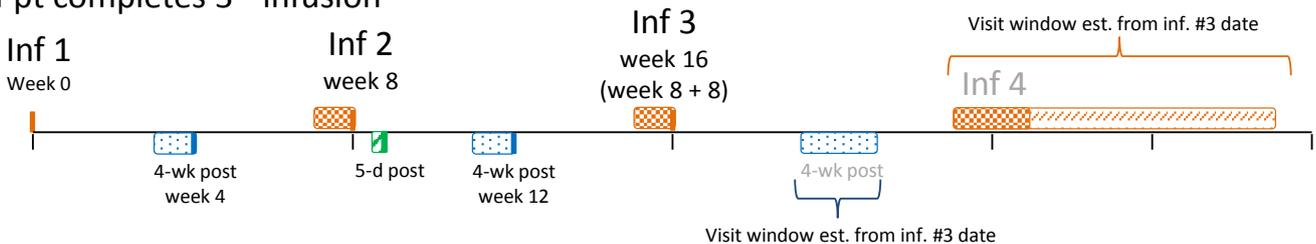

# Scenario 1:
## *Participant completes visits on target dates in middle of target windows*

A. Participant completes first infusion on day 0 of week 0. Two visit windows are created based on the date of the infusion: a 4-week post-infusion visit window centered on week 4 and an infusion visit window that allows the next infusion between 49 and 105 days (7 weeks – 15 weeks) later. Note that the target window is between day 49 and day 65 (first two weeks of window), but the database will allow an infusion visit through day 105.

B. Participant completes the 4-week post-infusion visit after the first infusion. The visit window for the 4-week post visit closes. The visit window for the second infusion is unaffected.

C. Participant completes the 2nd infusion on the target date (middle of window). The 2nd infusion window closes when the infusion is received. Being the second infusion, the completion of this infusion creates <u>three</u> new visit windows: a 5-day post infusion visit window, a 4-week post-infusion visit window centered on week 12, and the third infusion visit window 7-15 weeks after infusion #2.

D. Participant completes 5-day post and 4-week post visits after 2nd infusion. The visit windows for these visits close; the window for the 3rd infusion visit is unaffected.

E. Participant completes the third infusion, closing the infusion window. The 4-week post visit window and the window for the next infusion are created as described above.

# Scenario 2:
# Participant completes visits on dates toward end of target windows

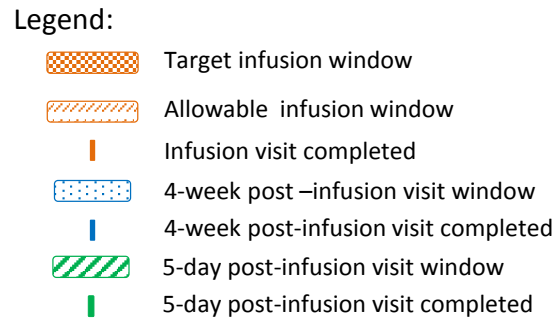

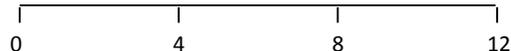

A. Ppt completes 1st infusion

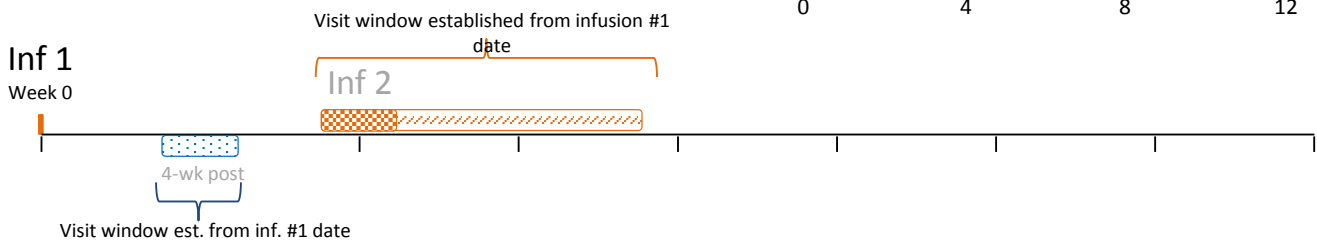

B. Ppt completes 4-week post after 1st infusion

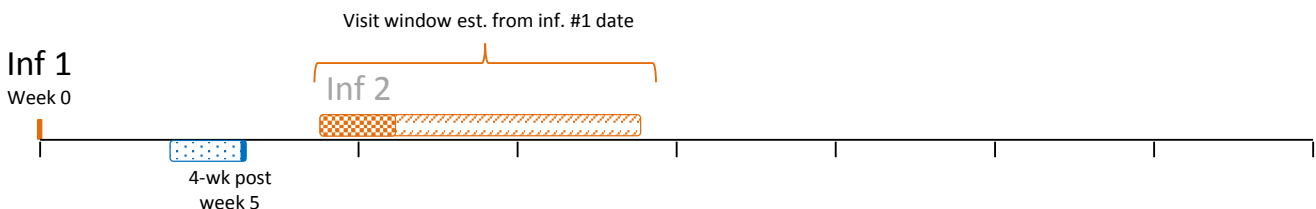

C. Ppt completes 2nd infusion

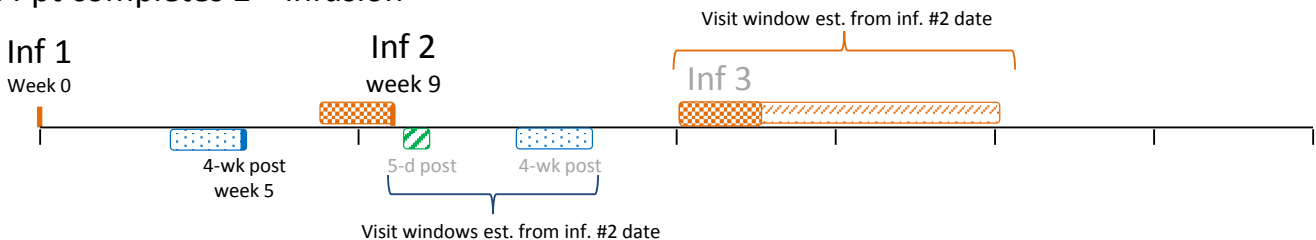

D. Ppt completes 4-week post after 2nd infusion

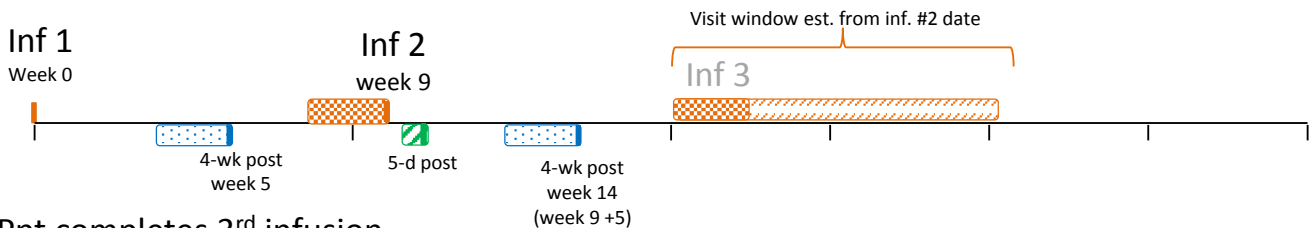

E. Ppt completes 3rd infusion

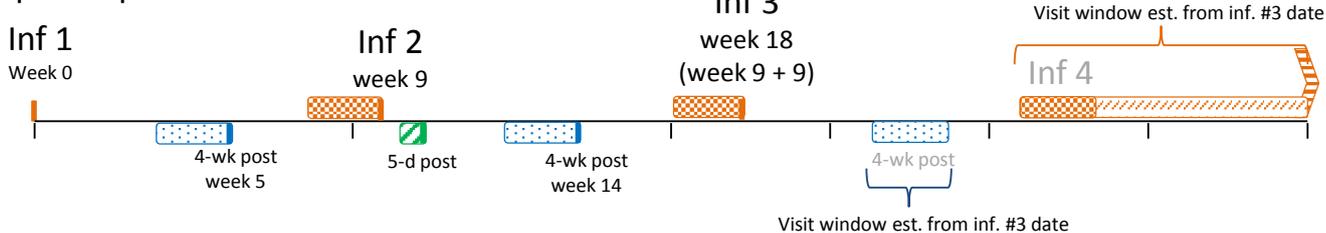

# Scenario 2:
## *Participant completes visits late in target windows*

A. Participant completes first infusion on day 0 of week 0. The 4-week post visit window and next infusion window are created as described above.

B. Participant completes the post-infusion visit late in the visit window. Even though this visit is late in the window, the next infusion window is unaffected since it is the prior infusion date that sets the target.

C. Participant completes the second infusion at the end of the target window. The 5-day and 4-week post-infusion visit windows and the window for infusion #3 are created based on the date of the second infusion.

D. Participant completes the 5-day post-infusion visit late in the visit window; this does not affect the 4-week post-infusion window. Participant completes the 4-week post-infusion visit late in the window, and as above, the next infusion visit window is unaffected.

E. Participant completes the third infusion late in the target window, and the subsequent 4-week post-infusion and next infusion visit windows are created based on the date of the third infusion.

# Scenario 3:
# Participant misses one infusion

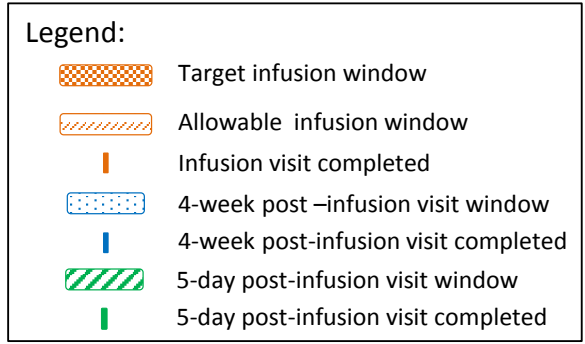

Legend:
- Target infusion window
- Allowable infusion window
- Infusion visit completed
- 4-week post –infusion visit window
- 4-week post-infusion visit completed
- 5-day post-infusion visit window
- 5-day post-infusion visit completed

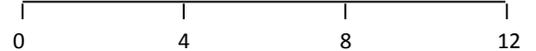

Scale (study weeks): 0, 4, 8, 12

### A. Ppt completes 1st infusion

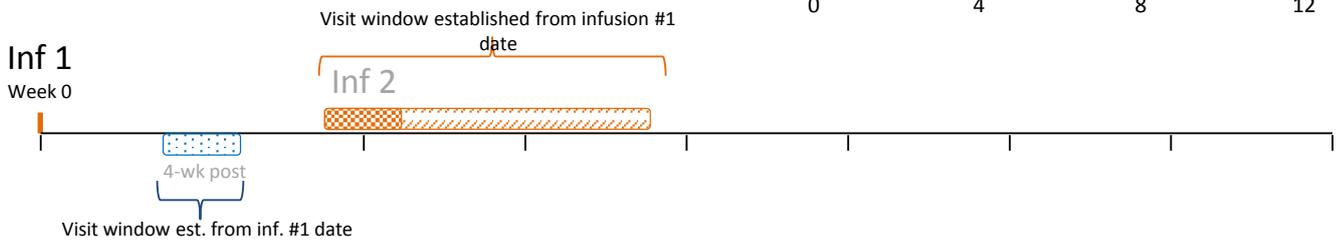

### B. Ppt completes 4-week post after 1st infusion

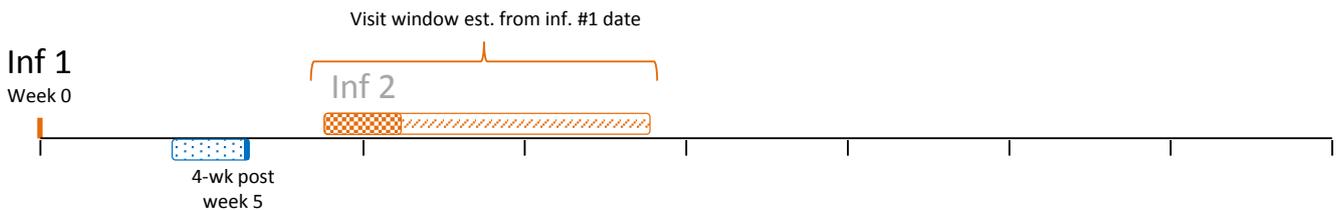

### C. Ppt misses 2nd infusion

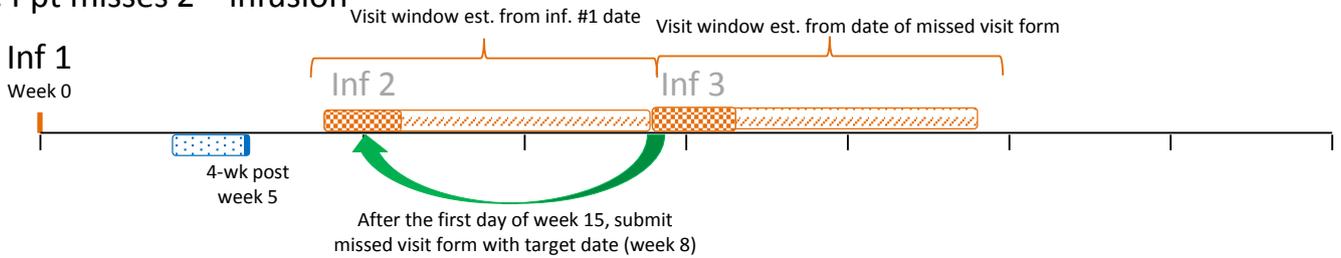

After the first day of week 15, submit missed visit form with target date (week 8)

### D. Ppt completes 3rd infusion

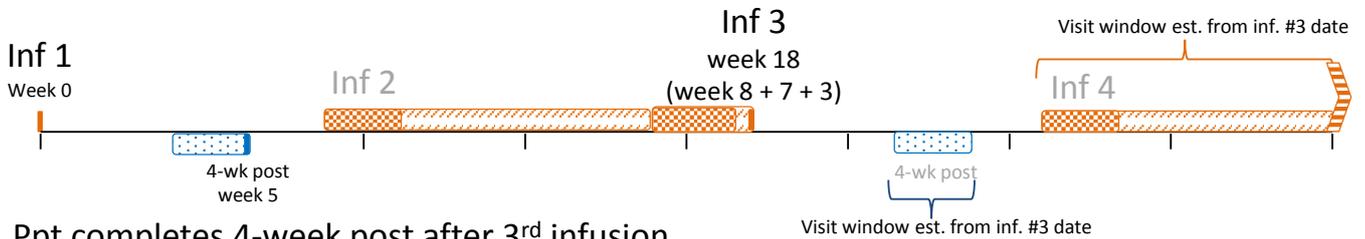

Inf 3 week 18 (week 8 + 7 + 3)

### E. Ppt completes 4-week post after 3rd infusion

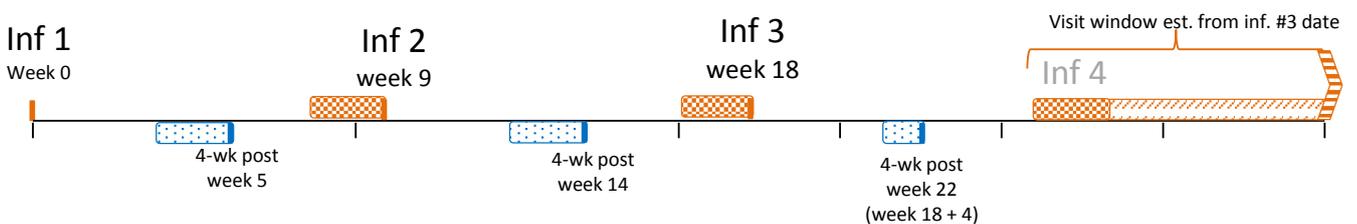

4-wk post week 22 (week 18 + 4)

# Scenario 3:
## *Participant misses one infusion*

A. Participant completes first infusion and subsequent windows are created as described above.
B. Participant completes the 4-week post-infusion visit.
C. The participant does not come into the clinic during the open window for infusion #2.  The window closes, and because there was no infusion, no 5-day or 4-week post-infusion visit windows were created. To create the next infusion window, the site submits a missed visit form with the missed visit date reflecting the target date for that infusion. Submission of the missed visit form results in the creation of the visit window for the next infusion, infusion #3.
D. The participant comes in to complete infusion #3 during the allowable window. This creates windows for both the 4-week post-infusion visit and the next infusion visit (#4) based on the date of infusion #3.
E. Participant completes the 4-week post-infusion visit from infusion #3 and the window for infusion #4 is unchanged.

# Scenario 4: Participant misses two infusions

Legend:
- Target infusion window
- Allowable infusion window
- Infusion visit completed
- 4-week post –infusion visit window
- 4-week post-infusion visit completed
- 5-day post-infusion visit window
- 5-day post-infusion visit completed

Scale (study weeks): 0, 4, 8, 12

### A. Ppt completes 1st infusion

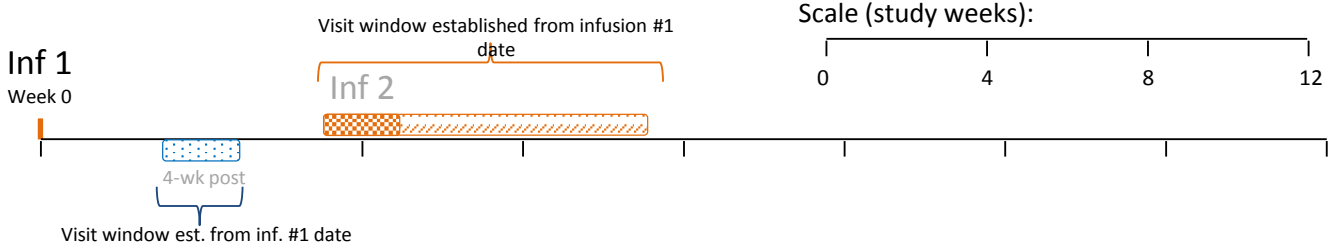

### B. Ppt completes 4-week post after 1st infusion

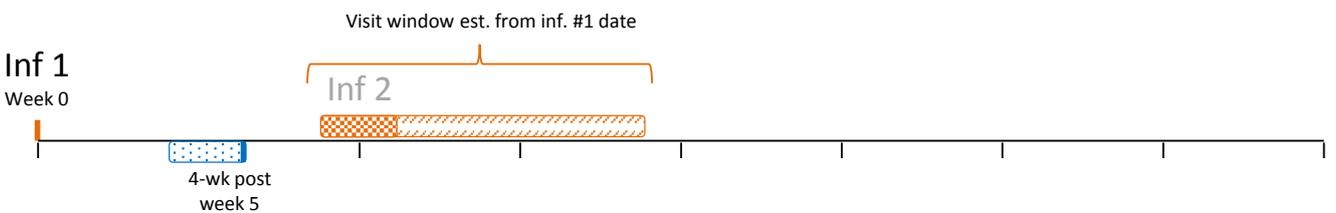

### C. Ppt misses 2nd infusion

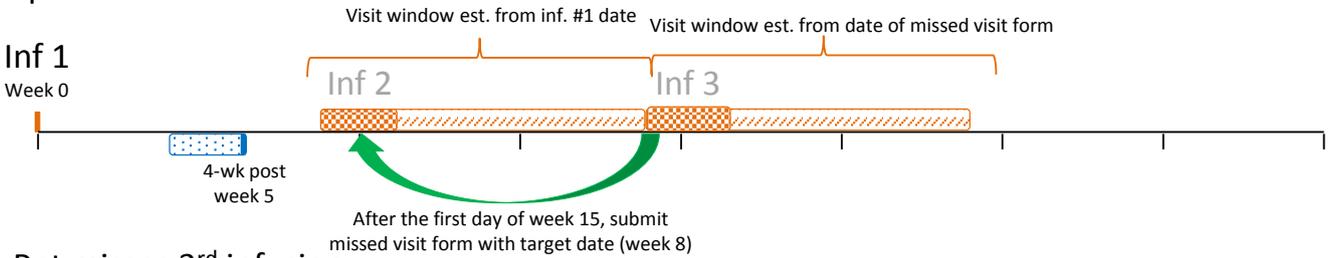

After the first day of week 15, submit missed visit form with target date (week 8)

### D. Ppt misses 3rd infusion

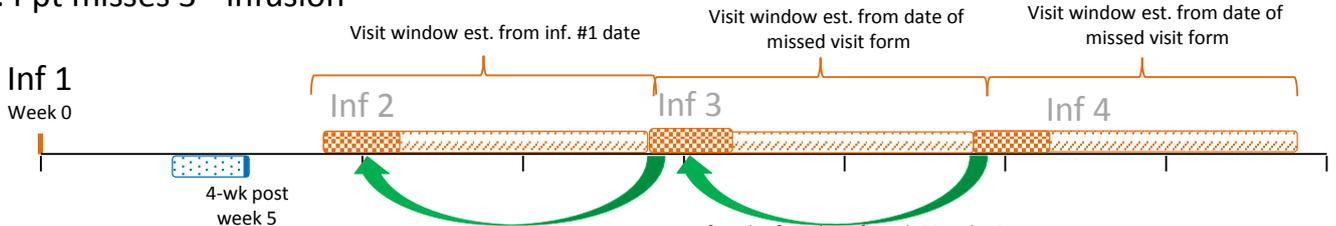

After the first day of week 23, submit missed visit form with target date (week 16)

### E. Ppt completes 4th infusion

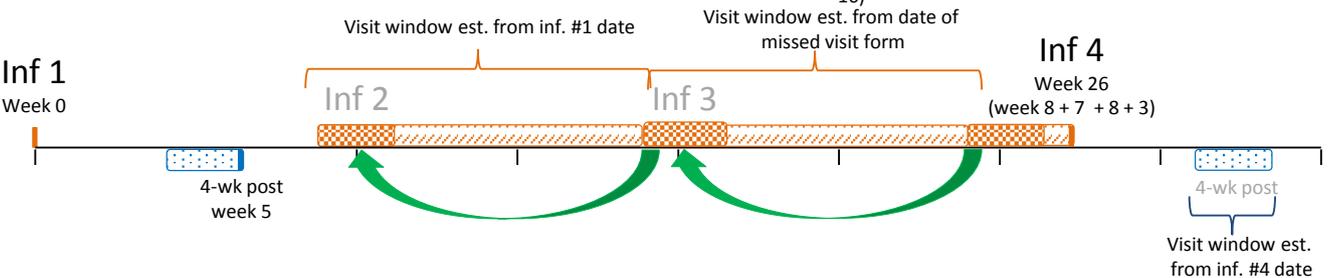

Inf 4 Week 26 (week 8 + 7 + 8 + 3)

Visit window est. from inf. #4 date

# Scenario 4:
## *Participant misses two infusions*

A. Participant completes first infusion and subsequent windows are created as described above.

B. Participant completes the 4-week post-infusion visit.

C. The participant does not come into the clinic during the open window for infusion #2. The window closes, and because there was no infusion, no 5-day or 4-week post-infusion visit windows were created. The site submits a missed visit form to create the visit window for the next infusion, infusion #3.

D. The participant does not come into the clinic during the open window for infusion #3. Because there was no infusion, no 4-week post-infusion visit window was created. The window closes, and the site submits a missed visit form to create the visit window for the next infusion, infusion #4.

E. Participant completes 4th infusion in the allowable window. This results in creation of the 4-week post-infusion visit and the window for the next infusion visit (off page).


\end{center}
\end{document}